\documentclass[12pt]{article}
\usepackage{latexsym}

\newcommand{\be}{\begin{equation}}
\newcommand{\ee}{\end{equation}}
\newcommand{\bea}{\begin{eqnarray}}
\newcommand{\eea}{\end{eqnarray}}
\newcommand{\bean}{\begin{eqnarray*}}
\newcommand{\eean}{\end{eqnarray*}}
\newcommand{\brray}{\begin{array}}
\newcommand{\erray}{\end{array}}
\newcommand{\ben}{\begin{equation}{nonumber}}
\newcommand{\een}{\end{equation}{nonumber}}
\newcommand{\newsection}[1]{\setcounter{dfn}{0}
\section{#1}}

\newtheorem{dfn}{Definition}[section]
\newtheorem{thm}[dfn]{Theorem}
\newtheorem{lmma}[dfn]{Lemma}
\newtheorem{ppsn}[dfn]{Proposition}
\newtheorem{crlre}[dfn]{Corollary}
\newtheorem{xmpl}[dfn]{Example}
\newtheorem{rmrk}[dfn]{Remark}

\newcommand{\bdfn}{\begin{dfn}}
\newcommand{\bthm}{\begin{thm}}
\newcommand{\blmma}{\begin{lmma}}
\newcommand{\bppsn}{\begin{ppsn}}
\newcommand{\bcrlre}{\begin{crlre}}
\newcommand{\bxmpl}{\begin{xmpl}}
\newcommand{\brmrk}{\begin{rmrk}}

\newcommand{\edfn}{\end{dfn}}
\newcommand{\ethm}{\end{thm}}
\newcommand{\elmma}{\end{lmma}}
\newcommand{\eppsn}{\end{ppsn}}
\newcommand{\ecrlre}{\end{crlre}}
\newcommand{\exmpl}{\end{xmpl}}
\newcommand{\ermrk}{\end{rmrk}}

\newcommand{\al}{\alpha}

\newcommand{\dlt}{\delta}

\newcommand{\del}{\partial}

\newcommand{\cla}{{\cal A}}
\newcommand{\clb}{{\cal B}}
\newcommand{\clc}{{\cal C}}
\newcommand{\cld}{{\cal D}}

\newcommand{\clh}{{\cal H}}
\newcommand{\cli}{{\cal I}}
\newcommand{\clj}{{\cal J}}
\newcommand{\clk}{{\cal K}}
\newcommand{\cll}{{\cal L}}

\newcommand{\cls}{{\cal S}}

\newcommand{\clz}{{\cal Z}}

\def\a*{{\cal A}_{h,*}}
\def\B{{\cal B}(h)}
\def\B1{{\cal B}_1(h)}
\def\b{{\cal B}^{s. a. }(h)}
\def\b1{{\cal B}^{s. a. }_1(h)}

\def\A{{\cal A}_\infty}
\def\h{h_\infty}

\newcommand{\itt}{\int \limits}

\newcommand{\ot}{\otimes}

\newcommand{\raro}{\rightarrow}

\newcommand{\lgl}{\langle}
\newcommand{\rgl}{\rangle}

\def \qed { \mbox{}\hfill $\Box$\vspace{1ex}}

\begin{document}
\begin{center}
{\Large {\bf   Stochastic Dilation  of Symmetric Completely Positive  Semigroups}}\\
\vspace{.2 in}
by\\
{\bf Debashish Goswami}\\
(Inst. F{\"u}r Angew. Math.,Wegelerstr. 6, Bonn),\\
{\bf Kalyan B. Sinha}\\
(Indian Statistical Institute, 203, B. T. Road, Kolkata-35, India).
\end{center}

\begin{abstract}
This is a continuation of the study of the theory of quantum
stochastic dilation of completely positive semigroups on a von
Neumann or $C^*$ algebra, here with unbounded generators. The
additional assumption of symmetry with respect to a semifinite
trace allows the use of the Hilbert space techniques, while the
covariance gives rise to better handle on domains. An Evans-Hudson
flow is obtained, dilating the given semigroup.
\end{abstract}
{\bf AMS Sub. Classification no.} : 81S25, 46L53
\section{Introduction}
In an earlier series of papers (\cite{GS}, \cite{GSP}), we had
constructed a theory of stochastic dilation ``naturally"
associated with a given completely positive (CP) semigroup (heat
semigroup) on a von Neumann or $C^*$ algebra with bounded
generator. There the computations involved $C^*$ or von Neumann
 Hilbert modules, using the results of \cite{CE}, map-valued
 quantum stochastic processes on modules and stochastic
 integration w.r.t. them (\cite{HP}, \cite{Par}). It is then
 natural to consider the case of a CP semigroup with unbounded
 generator and ask the same questions about the associated
 stochastic dilations. As one would expect, the problem is too
  intractable in this generality and we impose some further
  structures on it, viz. we assume that the semigroup is symmetric
  w.r.t. a semifinite trace and covariant under the action of a
  Lie group on the algebra. This additional hypothesis enables us
  to control the domains of the various operator coefficients
  appearing in the quantum stochastic differential equations so
  that the Mohari-Sinha conditions (\cite{MoS}) can be applied.
 The covariance is exploited again as in \cite{CGS} along with the
 assumption that the crossed-product von Neumann algebra $\cla
 >\!\!\!\!\lhd G$ is isomorphic with the von Neumann algebra generated by
 $\cla$ and the representation $u_g$  of $G$ in the GNS Hilbert space
   associated with the trace, to obtain the structure maps,  and
   finally the Evans-Hudson (E-H) flow is constructed essentially
   along the lines of the proofs in \cite{GS}. As precursors of
   this work, we may mention those in \cite{FaS} and \cite{AK}.
   While the first one deals with a general E-H flow with
   unbounded structure maps under some additional hypotheses, the
   second one treats the problem in a different spirit.
  \section{Preliminaries}
Let $\cla$ be a separable  $C^*$-algebra and $\tau$ be a densely defined, semifinite,
 lower semicontinuous and faithful  trace  on
 $\cla$.  Let  $\cla_\tau \equiv \{ x: \tau(x^*x)<\infty \}$.
   Let $h=L^2(\tau)$, and $\cla$ is naturally
 imbedded in $\clb(h)$. We denote by $\bar{\cla}$ the von Neumann closure of $\cla$ with respect to the weak
  topology inherited from $\clb(h)$. Clearly $\cla_\tau$ is ultraweakly dense in $\bar{\cla}$.
 Assume furthermore that $G$ is a second countable Lie group with $( \chi_i,i=1,...N )$  a basis of its
  Lie algebra, $g \mapsto \alpha_g \in Aut(\cla)$  a strongly continuous representation. Suppose that
  $\al_g(\cla_\tau)\subseteq \cla_\tau $ and $\tau(\al_g(x^*y))=\tau(x^*y)$ for $x \in \cla_\tau, y\in \cla$, $g \in G$
   (by polarization this is equivalent to the
     assumption that $\tau(\al_g(x^*x))=\tau(x^*x)$ for $x \in \cla_\tau$ ). This allows one to extend $\al_g$ as
    a unitary linear operator (to be denoted by $u_g$) on $h$ and clearly $\al_g(x)=u_g x u_g^{*}$ for $x \in \cla$. It is
     indeed easy to verify this relation on vectors in $\cla_\tau$ and then it extends to the whole of $h$ by the fact that
      $h$ is the completion of $\cla_\tau$.
   For $f \in C_c^\infty(G)$ (i.e. $f$ is smooth complex-valued function with compact support on $G$) and
 an element $x \in \cla$, let us denote by
 $\al(f)(x)$ the norm-convergent integral $\int_G f(g) \al_g(x)dg$, where $dg$ denotes the
 left Haar measure on $G$.
   \blmma
         $g \mapsto u_g$ is  strongly continuous w.r.t the Hilbert-space topology of
       $h$.
             \elmma
         {\it Proof :-}\\
         Let $\cla_1 \equiv \{ x \in \cla | \tau(|x|) < \infty
         \}$. It is known that $\cla_1$ is dense in $h$ in the
          topology of $h$. Furthermore, for $x \in \cla_\tau$ and $y \in \cla_1$,
           $|\tau((u_g(x)-x)^*y)| \leq
           \|(u_g(x)-x)^*\|\tau(|y|)$, which proves that $g
           \mapsto \tau((\al_g(x)-x)^*y)$ is continuous, by the
           strong continuity of $\al$ w.r.t. the norm topology of
           $\cla$. But by the density of $\cla_1$ and $\cla_\tau$
           in $h$ and the fact that $u_g$ is unitary, we conclude
           that for fixed $\xi \in h$, $g \mapsto u_g \xi$ is
           continuous w.r.t. the weak topology of $h$, and hence
           is strongly continuous.
           \qed

The above lemma allows us to define $\al(f)(\xi)=\int f(g) u_g
(\xi) dg \in h$ for $f \in C_c^\infty(G), \xi \in h$.
     Furthermore,     from the  expression  $\al_g(x)=u_gxu_g^*$,
 it is possible to extend $\al_g$ to the whole
       of $\clb(h)$ as a normal automorphism group implemented by the unitary group $u_g$ on $h$ and we shall denote
       this extended automorphism group too by the same notation. Let $\A \equiv \{x \in \cla : g \mapsto \al_g(x)$  is
         infinitely
        differentiable  w.r.t. the  norm  topology $\}$, i.e. $\A$ is the intersection of the domains of
        $\del_{i_1} \del_{i_2}
         ... \del_{i_k}; k \geq 1$, for all possible $i_1,i_2,...\in \{1,2,...N \}$, where $\del_i$ denotes the closed $\ast$-derivation
          on $\cla$   given by the generator  of the one-parameter automorphism group
          $\al_{exp(t\chi_i)}$, where $exp$ denotes
           the usual exponential map for the Lie group $G$.
         The following result  is essentially a consequence of   the results obtained in \cite{Gar},
 \cite{Nel}.
      \bppsn
 \label{sobo}
  (i) $\A$ is dense  $\ast$-subalgebra of $\cla$.\\
 (ii) Similarly, we denote by $d_k$ the self-adjoint
               generator of the unitary group $u_{exp(t \chi_k)}$ on $h$ such that
               $u_{exp (t \chi_k)}=e^{it d_k}$, and
                let $\h \equiv \bigcap_{i_1,i_2,...} Dom(d_{i_1} d_{i_2}...d_{i_k};k=1,2,...)$.
        Then $\h$ is dense in $h$.\\
    (iii)       If we equip $\A$ with a family of norms $\| . \|_{\infty,n}; n=0,1,2,...$
                    given by :
                 $$
                 \| x \|_{\infty,n}=\sum_{i_1,i_2,...i_k;k \leq n } \|\del_{i_1}...\del_{i_k} (x)\|;
                 $$
                 for $n \geq 1$, and $ \| x \|_{\infty,0}=\|x\|$,
        and   similarly  define a family of Hilbertian norms
                  $\| . \|_{2,n}; n=0,1,2,...$ on $\h$ by :\\
                  $$ \| \xi\|_{2,n}^2\equiv \sum_{i_1,i_2,...i_k;k\leq n} \| d_{i_1}d_{i_2}
                  ...d_{i_k}(\xi)\|^2 $$ on $h_\infty$, then
                          $\A$ and $\h$ are complete with respect to
   the locally convex  topologies induced by
  the respective (countable) family of norms as defined above. In other words, $\A$ and
 $h_\infty$ are Frechet spaces in the topologies (to be called ``Frechet topologies" from now on) described above.\\
     (iv) $\al_g(\A) \subseteq \A$, $u_g(h_\infty) \subseteq h_\infty$ for all $g \in G$. Furthermore,
       $g \mapsto \al_g(x), g \mapsto u_g(\xi)$ are smooth ($C^\infty$) in the respective Frechet topologies for
 $x \in \A, \xi \in h_\infty$.\\
(v) Let $\cla_{\infty, \tau}= \cla_\infty \bigcap \h  $. It is a $\ast$-closed two-sided ideal in
  $\A$ and
    is dense in   $\cla$, $\A$, $h$ and  $\h$ w.r.t. the relevant topologies.
\eppsn
                          Proof : \\
  The proof of (i) and (ii)  will follow immediately from the references cited before the statement of this
 proposition.
 The proof of (iii) is quite standard, which uses the fact that $\del_i,d_i$'s are closed maps in
 $\cla$ and $h$ respectively.

  Next we indicate briefly the proof of (iv) for $\A$ only, since it is similar for
 $h_\infty$. First of all, by the definition of $\A$ and the fact that $G \times G \ni (g_1,g_2)
 \mapsto g_1g_2 \in G$ is $C^\infty$ map, we observe that for $x \in \A$
  the map $(g_1,g) \mapsto \al_{g_1}(\al_g(x))=\al_{g_1g}(x)$ is $C^\infty$ on
 $G \times G$, hence in particular for fixed $g$, $G \ni g_1 \mapsto \al_{g_1}(\al_g(x))$ is
 $C^\infty$, i.e. $\al_g(x) \in \A$. Similarly,
  for fixed $x \in \A$ and any positive integer $k$, the map $F : R^k \times G \raro \cla$
 given by $F(t_1,...t_k,g)=\al_{exp(t_1\chi_{i_1})...exp(t_{i_k} \chi_k)g}(x)$
 is $C^\infty$.
 By  differentiating $F$ in its first $k$ components
 at $0$, we get that $\partial_{i_1}...\partial_{i_k}(\al_g(x))$
 is $C^\infty$ in $g$.

  To prove (v), we need to  note first that the  elements of the form $\al(f)(\xi)$, with
 $f \in C_c^\infty(G)$ and $\xi \in \cla_\tau$ are  clearly in $\cla_{\infty, \tau}$.
 Let us first consider the density in  $h$ and $\h$.  Since the topology of $h_\infty$ is stronger
 than that of $h$ and since $\h$ is dense in $h$ in the topology of $h$, it suffices to prove
 that the set of elements of the above form  is dense in $h_\infty$ in the Frechet topology.
  For this, we take $\xi \in \h$, and choose a net
  $x_\nu$ of elements from $\cla_\tau$ which converges in the topology of
 the Hilbert space $h$ to $\xi$, and then it is clear that $\al(f)(x_\nu) \raro
 \al(f)(\xi)$ $\forall f \in C_c^\infty(G)$ w.r.t. the Frechet topology of $\h$,
 since $d_{i_1}...d_{i_k}\al(f)(x_\nu-\xi)
=(-1)^k\al(X_{i_1}...X_{i_k}f)(x_\nu-\xi)$. Thus, it is enough to show that
 $\{ \al(f)(\xi), f \in C_c^\infty(G), \xi \in \h \}$ is dense in $\h$ in the Frechet
 topology. For this, we  choose a net $f_p \in C_c^\infty(G)$
 such that $\int_G f_p dg =1 \forall p$ and the support of $f_p$ converges to the singleton
 set containing the identity element of the group $G$, and then it is simple to see that
 $\al(f_p)(\xi) \raro \xi$ in the Frechet topology.     Finally, the norm-density
 of $\cla_{\infty,\tau}$ in $\cla$ and the Frechet density in $\A$ will follow by similar arguments.
                        \qed\\

\brmrk
It may be noted that for $x \in \cla_{\infty,\tau}$, $\delta_{i_1}...\delta_{i_k}(x)=
 d_{i_1}...d_{i_k}(x) \in \cla \bigcap h.$ This follows from the  fact that if
 $y_p$ is a net in $\cla \bigcap h$ which converges both in the norm topology of $\cla$ as well
 as in the Hilbert space topology of $h$, then the norm-limit belongs to $h$ and
 the  two limits must coincide as vectors of $h$.
\ermrk


    Now we shall introduce some more useful notation and terminology and prove some preparatory results. If $\clh$ is any Hilbert
     space with a strongly continuous unitary representation of $G$ given by $U_g$, we denote by $\clh_\infty$ the intersection
      of the domains of the self-adjoint generators of different one-parameter subgroups, just as we did in case of $h$.
       We denote the corresponding family of ``Sobolev-like" norms again by the same notation as in case of $h$ and consider
        $\clh_\infty$ as a Frechet space as earlier.
         We call such a pair $(\clh, U_g)$ a Sobolev-Hilbert space and for  two such
         pairs $(\clh,U_g)$ and $(\clk, V_g)$, we denote by $\clb(\clh_\infty, \clk_\infty) $ the space of all linear maps $S$ from
          $\clh$ to $\clk$  such that  $\clh_\infty$ is in the domain of $S$,
          $S(\clh_\infty)\subseteq \clk_\infty$,   and $S$ is continuous with respect to the Frechet topologies of the respective spaces.
           We call a linear map $L$ from $\clh$ to $\clk$ to be {\bf covariant} if
           $\clh_\infty \subseteq Dom(L)$ and $LU_g(\xi)=V_g L(\xi)  \forall g \in G,
            \xi \in \clh_\infty$.
        \blmma
        \label{bddcov}
        If $L$ from $\clh$ to $\clk$ is bounded (in the usual Hilbert space sense) and covariant in the above sense, then
         $L \in \clb(\clh_\infty, \clk_\infty)$.
         \elmma
         Proof :\\
         Let $d_i^\clh$ and $d_i^\clk$ be respectively the self-adjoint generator of the one parameter subgroup
          corresponding to $\chi_i$ in $\clh$ and $\clk$. From the relation $LU_g=V_g L$ it follows that
           (since $L$ is bounded) $L$ maps the domain of $d_i^\clh$ into the domain of $d_i^\clk$ and $L d_i^\clh=
            d_i^\clk L$. By repeated application of this argument it follows that $L d_{i_1}^\clh ... d_{i_k}^\clh (\xi)
             =d^\clk_{i_1}...d_{i_k}^\clk L (\xi) \forall \xi \in \clh_\infty$, and thus $\| L \xi \|_{2,n}\leq
              \| L\| \| \xi \|_{2,n}$.
              \qed

              We shall call an element of $\clb(\clh_\infty, \clk_\infty)$ a ``{\bf smooth}" map, and if such a smooth map
               $L$ satisfies an estimate $\| L \xi \|_{2,n} \leq C \| \xi \|_{2,n+p}$ for all $n$ and
                for some integer $p$ and a constant $C$,
                then we say that $L$ is a smooth map of order $p$ with the bound $\leq C$. From the proof of the above lemma
                 we observe that any bounded covariant map is smooth of order $0$ with the
                 bound $\leq \| L\|$. By a similar reasoning
                  we can prove the following :\\
                  \blmma
                  \label{cov}
                   Suppose that $L$ is a closed (in the Hilbert space sense),
                    covariant map from $\clh$ to
                    $\clk$ and $\clh_\infty$ is in the domain of $L$.
                      Under these assumptions, $L$ is smooth of the order $p$ for some $p$.
                  \elmma
                  Proof :\\
                  For simplicity of notation, we shall use the same symbol $d_i$ for both $d_i^\clh$
                   and $d_i^\clk$, and also we use the same symbols for the corresponding one parameter
                    groups of unitaries acting on $\clh$ and $\clk$.
                    Let $L$ be a map as above. Since $L$ is closed in the Hilbert space sense,
                     and the Frechet topology in $\clh_\infty$ is stronger than its
                      Hilbert space topology, it follows that $L$ is closed as a map
                       from the Frechet space $\clh_\infty$ to the Hilbert space $\clk$, and
                        being defined on the entire $\clh_\infty$, it is continuous w.r.t the above
                         topologies. By the definition of Frechet space continuity,
                     there exists some $C$ and $p$ such that $\|L(\xi)\|_{2,0}\leq C \| \xi\|_{2,p}$.
                     Now, for any fixed $k$, let $u_t\equiv u_{exp(t \chi_k)}$.  Since $u_t$ maps
                      $\h$ into itself and $L$ is covariant, we have that $L(\frac{u_t(\xi)-\xi}{t})=
                       \frac{u_t(L\xi)-L \xi}{t}$. Now, since  $\frac{u_t(\xi)-\xi}{it} \raro d_k(\xi)$
                 as $t \raro 0+$ in the Frechet topology, we have
                 that
                               $L(\frac{u_t(\xi)-\xi}
         {it})=\frac{u_t(L\xi)-L\xi}{it}$ converges to $Ld_k\xi$ in the
                          Hilbert space topology of
                                 $\clk$, and so by the closedness of $d_k$
                                   $L\xi$ must belong to the domain of $d_k$, with $Ld_k \xi=
                                 d_kL\xi$. Repeated use of this argument proves that
                         $L(\clh_\infty)\subseteq \clk_\infty$ and $L(d_{i_1}...d_{i_k}\xi)
       =d_{i_1}...d_{i_k}(L \xi) \forall \xi \in \clh_\infty$.
                          Now, a direct computation enables one to show that $L$ is of order $p$
                              with the bound $\leq C$.

                              \qed

                 \bthm
                 \label{*cont}
       Let $(\clh,U_g),(\clk,V_g)$ be two Sobolev-Hilbert spaces as in earlier discussion,
        and $L$ be a closed (not as Frechet space map but as Hilbert space map)
        linear map from  $ \clh$ to $\clk$. Furthermore, assume that
        $\clh_\infty$ is in the domain of $|L|^2$
      and is         a core for  $|L|^2$, and $LU_g=V_gL$ on $\clh_\infty$.
            Then we have the following conclusions :\\
          (i) $L$ is a smooth covariant map with some order $p$ and  bound $\leq C$ for some $C$;\\
          (ii) $L^*$ (the densely defined adjoint in the Hilbert space sense ) will have $\clk_\infty$
           in its domain;\\
           (iii) $L^* $ is also  a smooth covariant map from $\clk_\infty$ to $\clh_\infty$; with
            order $p$ and bound $\leq C$ as in (i). \\

           \ethm
        Proof :\\

                  Let the polar decomposition of $L$ be given by $L=W|L|$. We claim that
                  both $W$ and $|L|$ are covariant maps. First we note that  $\clh_\infty$
                   is also a core for $L$ (being a core for $|L|^2$) and since $U_g$ is a unitary operator
                    that maps $\clh_\infty$ into itself, clearly $\clh_\infty$ is a core for $LU_g$ and also
                     for $V_gL$. Thus the relation $LU_g=V_gL$ on $\clh_\infty$ implies that the operators
                      $LU_g$ and $V_gL$ have the same domain and they are equal.
                       Now, note that $L$ being closed and $V_g$ being bounded, we have that
                        $(V_gL)^*=L^*V_g^*=L^*V_{g^{-1}}$. Furthermore, since $U_g^{-1}$ maps the core
                         $\clh_\infty$ for $L$ into itself, one can easily verify that $(LU_g)^*=U_g^*L^*$
                                Thus,  we get that $U_gL^*=L^*V_g \forall g$. It then  follows that
                         $U_g|L|^2=|L|^2U_g$ and hence by spectral theorem $U_g $ and $|L|$ will commute.
                         By Lemma \ref{cov}, we get that
                                     $|L|(\clh_\infty)\subseteq \clh_\infty$,
      and $|L|$ is a smooth covariant map of some order.

                           Now, if $P$ denotes the projection
                           onto the closure of the range of $|L|$, then $P$ clearly commutes with
                            $U_g$ for all $g$, hence in particular $U_gRan(P)^\perp \subseteq Ran(P)^\perp$. Thus
                             $WU_gP^\perp=WP^\perp U_g=0=V_gWP^\perp$. On the other hand, $V_gWP=WU_gP$, because
                              $V_gW|L|=V_gL=LU_g=W|L|U_g=WU_g|L|$. Hence we have that $W$ is a bounded covariant map,
                               and thus by \ref{bddcov}, it follows that $W^*$ is covariant too,
                                and in particular $W^*(\clk_\infty)
                               \subseteq \clh_\infty \subseteq Dom(|L|)$, so that $\clk_\infty \subseteq
                                Dom(L^*)=Dom(|L|W^*)$. Furthermore, from the fact that $W$ and $W^*$ are
                                 smooth maps of order
                                 $0$ with bound $\leq 1$ (as $\|W\|=\|W^*\|=1$) and $|L|$ is a smooth covariant map
                                  of some order $p$
                                 with  bound $\leq C$ for some $C$, clearly both $L=W|L|$ and $L^*=|L|W^*$
                                  are smooth covariant maps of order $p$ and bound $\leq C$, which completes the proof.
                                  \qed

            \blmma
            \label{tensor}
            Let $(\clh_i,U^i_g),i=1,2$ and $(\clk_i,V^i_g),i=1,2$ be Sobolev Hilbert spaces
             and $k$ be any
             Hilbert space. Then we can construct Sobolev Hilbert spaces $(\clh_i \oplus \clk_i,U^i_g \oplus
              V^i_g)$ and $(\clh_i \otimes k, U^i_g \otimes I$ (with the symbols carrying their usual meanings)
               and  if $L \in \clb(\clh_{1_\infty}, \clh_{2_\infty}), M\in \clb(\clk_{1_\infty},
                \clk_{2_\infty})$, then we have the following :\\
                (i)  $L \oplus M $ is a smooth map between appropriate spaces, and \\
                (ii) $(\clh_1 \ot k)_{\infty}$ is the completion of $\clh_{1_\infty}\ot_{alg}k$
                 under the respective Frechet topology and the map $L \otimes_{alg} I $ on $\clh_{1_\infty} \otimes_{alg} k $
                 extends as a smooth map on the respective Frechet space (we shall denote this smooth
                  map by $L \ot I$ or sometimes $\tilde{L}$). Furthermore, if $L$ is of order
                  $p$ with some constant $C$, so will be $\tilde{L}$.
                  \elmma
                  Proof :\\
                  (i) is straightforward. To prove (ii), we fix any orthonormal basis $\{e_l\}$ of
                   $k$ and let $\xi=\sum \xi_l \ot e_l$ be a vector in the domain of the self adjoint
                   generator of the one parameter unitary group $u_t \ot I$, where
                                       $u_t$ is as in the proof of Lemma \ref{cov} and  the summation is over a countable set since
 $\xi_l=0$ for all but countably many values of $l$. So, without loss of generality we may assume
  that the set of $l$'s with $\xi_l$ nonzero is indexed by $1,2,...$.   Since $\sum (\frac{u_t(\xi_l)-\xi_l}{t})\ot e_l$
                    is Cauchy (in the Hilbert space topology )
                    suppose that $\sum (\frac{u_t(\xi_l)-\xi_l}{t})\ot e_l \raro \sum \eta_l \ot e_l$.
                     Clearly, for each $l$, $\eta_l=\lim_{t \raro 0}(\frac{u_t(\xi_l)-\xi_l}{t})$, which implies that
                      $\xi_l \in Dom(d_k)$ and $d_k \xi_l=\eta_l$.  Thus, if $\tilde{d_k}$ denotes the self adjoint generator of
                         the one parameter unitary group $u_t \ot I$, then we have  proved that the domain of it
                          consists   of precisely the vectors $\sum \xi_l \ot e_l$ such that each $\xi_l \in
                           Dom(d_k)$ and $\sum \| d_k(\xi_l)\|^2<\infty$. Repeated use of this argument enables us
                            to prove that $(\clh_1 \ot k)_\infty$ consists of the vectors $\xi=\sum \xi_l \ot e_l$
                             with the property that $\xi_l \in \clh_{1_\infty} \forall l$ and for any $n$,
                              $\|\xi\|_{2,n}^2 \equiv \sum_l \| \xi_l  \|_{2,n}^2 < \infty$.
                               From this, it is clear that $\sum_{l=1}^m \xi_l \ot e_l$
                                converges (as $m\raro \infty$) to $\xi$ in each of the $\|.\|_{2,n}$ norms, i.e. in the
                                 Frechet topology. The rest of the proof follows by observing that
                                  for any $\xi=\sum_{\rm finite} \xi_l \ot e_l \in \clh_{1,\infty} \ot_{\rm alg}
                                   k,$ $\|\tilde{L}(\xi)\|_{2,n}^2=\sum \|L\xi_l\|_{2,n}^2$. \\
                              \qed

 \section{Assumptions on the semigroup and its generator}

 Let $T_t$ be a q.d.s. on $\cla$ which is $\tau$-symmetric (i.e. $\tau(T_t(x)y)=
 \tau(xT_t(y))$ for all positive $x,y \in \cla$, and for all $t \geq 0$). We refer the reader
 to \cite{CiS} for a detailed account of such semigroups from the point of view of
 Dirichlet forms. We shall need some of the results obtained in that reference. As it is
 mentioned in that reference, $T_t$ can be canonically extended to a normal
  $\tau$-symmetric q.d.s. on $\bar{\cla}$ as well as to $C_0$-semigroup of
  positive contractions on the Hilbert space $h$. We shall denote all these semigroups
 by the same symbol $T_t$ as long as no confusion can arise. Furthermore, we assume that
 $T_t$ on $\bar{\cla}$ is conservative, i.e. $T_t(1)=1 \forall t \geq 0$.

 Let us denote by $\cll$ the $C^*$ generator of $T_t$ on $\cla$, and by $\cll_2$ the
 generator of $T_t$ on $h$. Clearly, $\cll_2$ is a negative self-adjoint map
 on $h$. We also recall (\cite{CiS}) that there is a canonical Dirichlet form
  $\eta$ on $h$ given by, $Dom(\eta)=Dom((-\cll_2)^{\frac{1}{2}})$, $\eta(a)
 =\| (-\cll_2)^{\frac{1}{2}}(a) \|_{2,0}^2, a \in Dom(\eta)$. We
 recall from \cite{CiS} that $\clb:=\cla \bigcap Dom(\eta)$ is a
 $\ast$-algebra, called the Dirichlet algebra, which is norm-dense
 in $\cla$.

We now make the following important assumptions :\\
{\bf Assumptions:}\\
 ({\bf A1}) $T_t$ is covariant, i.e. $T_t$ commutes with $\al_g$
 for all $t \geq 0, g \in G$.\\
({\bf A2})$\cll$ has $\A$ in its domain,\\
({\bf A3}) $\cll_2$ has  $h_\infty$ in its domain. \\

 The assumption that $T_t$ is covariant in particular implies
   that $T_t$ leaves $\cla_\infty$ invariant, hence by Nelson's theorem
    this domain is a core for $\cll$, and clearly $\al_g \cll=\cll \al_g$ on $\cla_\infty$.
     It follows  that (by arguments similar
 to those in the proof of \ref{sobo})
  $\cll(\cla_\infty)\subseteq \cla_\infty$. Similarly, $\cll_2(\h) \subseteq
 \h$. Since the actions of $\cll$ and $\cll_2$ coincide on $\cla_{\infty,\tau}$,
 one has that $\cll(\cla_{\infty,\tau}) \subseteq \cla_{\infty,\tau}$.
 Furthermore, we have the following :\\

\blmma
\label{core}
 $\cla_{\infty,\tau}$  is stable under the action of $T_t$ and  hence
   is a core for both $\cll_2$ and $\cll$.
\elmma
The proof of the lemma is straightforward and hence omitted.\\

By the Lemma \ref{core}, $\h$ is also a  core for $\cll_2$, as $\cla_{\infty,\tau}
 \subseteq \h$. It is important to remark here that the assumption {\bf A3} is the only hypothesis
  on the generator of the semigroup which involves the generator at the $L^2$-level, not
 the norm generator. However, we shall later on see that in an important special case, where
 the group $G$ is compact and acts ergodically on the algebra $\cla$, the assumption {\bf A3} will
 follow automatically from the other hypotheses.

Modifying slightly the arguments of  \cite{CiS} and \cite{Sav}, we
describe the structure of  $\cll$.
\bthm
\label{sv}
 (i) There is a Hilbert space $\clk$ equipped with an
 $\cla$-$\cla$ bimodule structure. We denote the right action by
 $(a,\xi) \mapsto \xi a, \xi \in \clk, a \in \cla$ and the left
 action by  $(a,\xi) \mapsto \pi(a) \xi, \xi \in \clk, a \in
 \cla$. \\
 (ii) There is a densely defined closable linear map $\delta_0$
 from $\cla$ into $\clk$ such that $\cla_{\infty,\tau} \subseteq \clb =Dom(
 \delta_0)$ (where $\clb$ is the Dirichlet algebra mentioned
 earlier), and $\delta_0$ is a bimodule derivation, i.e.
 $\delta_0(ab)=\delta_0(a)b+\pi(a)\delta_0(b) \forall a,b \in
 \clb$. \\
 (iii) For $a,b \in \cla_{\infty,\tau}$, $\| \delta_0(a)b \|_\clk
 \leq C_a \| b \|_{2,0},$ where $\| . \|_\clk$ denotes the Hilbert
 space norm of $\clk$, and $C_a$ is a constant depending only on
 $a$. Thus, for any fixed $a \in \cla_{\infty,\tau}$, the map $\cla_{\infty,\tau} \ni b
 \mapsto \sqrt{2} \delta_0(a)b \in \clk$ extends to a unique bounded linear
 map between the Hilbert spaces $h$ and $\clk$, and this bounded
 map will be denoted  by $\delta(a)$. \\
 (iv) $$ \partial \cll (a,b,c) \equiv  \dlt(a)^* \pi(b) \dlt(c)  = \cll(a^*bc)-\cll(a^*b)c-a^*\cll(bc)+
 a^* \cll(b)c,$$
 for $a,b,c \in \cla_{\infty,\tau}$.\\
(v) $\clk$ is the closed linear span of the vectors of the form
 $\delta(a)b, a,b \in \cla_{\infty,\tau}$.\\
(vi)  $\pi$ extends as a  normal $\ast$-homomorphism on
$\bar{\cla}$.\\

 \ethm
{\it Proof :-}\\
 We refer for the proof of (i) and (ii) to  \cite{Sav} and \cite{CiS}. Now, we
   note that $\cla_{\infty,\tau}$ is contained in the
 ``Dirichlet algebra" (c.f. \cite{CiS}) and in fact is a form-core for the Dirichlet form
 $\eta$ mentioned earlier. Using the calculations made in the
 proof of Lemma 3.3 of [8, page 8], we see that for $a,b
 \in \cla_{\infty,\tau}$,
 $$ \| \delta_0(a)b \|_\clk^2 =\frac{1}{2}
 \tau(-b^*\cll(a)^*ab-b^*a^*\cll(a)b+b^*\cll(a^*a)b).$$
  Here, we have also used the fact that $a,a^*,a^*a \in
  Dom(\cll)$. From the above expression (iii) immediately follows.
 We verify (iv) by direct and straightforward calculations, which
 we omit. To prove (v), we first recall from \cite{CiS} that
 $\clk$ can be taken to be  the closed linear span of the vectors
 of the form $\delta_0(a)b, a,b \in \clb$. Now, by Lemma 3.3 of
 \cite{CiS}, $\| \delta_0(a)b\|_\clk^2 \leq \|b\|_{\infty,0}^2
 \eta(a,a)$. Since $\cla_{\infty,\tau}$ is on one hand norm-dense
 in $\cla$ and also form core for $\eta$ on the other hand, (v)
 follows.

 Let us now prove (vi). It is enough to prove that whenever we have a Cauchy net
  $a_\mu \in \cla_{\infty,\tau}$
 in the weak
 topology, then
  $\lgl \xi, \pi(a_\mu) \xi \rgl $ is also Cauchy for any fixed $\xi$ belonging to
 the dense subspace of $\clk$ spanned by the vectors of the form $\dlt(b)c, b,c \in \cla_{\infty,\tau}.$
 But it is clear that for this, it suffices to show that $a \mapsto
  \lgl \dlt(b)b^\prime, \pi(a) \dlt(b)b^\prime \rgl
  $ is weakly continuous.   Now, by the symmetry of $\cll$  and the trace property of $\tau$,
  we have that for $a \in \cla_{\infty,\tau},$
  $$ \lgl \dlt(b)b^\prime, \pi(a)\dlt(b)b^\prime \rgl =
   \lgl b,ab\cll(b^\prime {b^\prime}^*)\rgl -\lgl b, a \cll(bb^\prime {b^\prime}^*)\rgl
     -\lgl \cll(bb^\prime {b^\prime}^*), ab\rgl+
 \lgl \cll(bb^\prime (bb^\prime)^*),a \rgl.  $$
   The  first three terms in the right hand side are clearly weakly continuous in $a$,
     so we have to concentrate only on the last term, which is of
     the form $\tau(\cll(xx^*)a)$ for $x \in \cla_{\infty,\tau}.$
      Now, we have,
     $$\tau(\cll(xx^*)a)=\tau(\cll(x)x^*a)+\tau(x\cll(x^*)a)+\tau(\delta(x^*)^*\delta(x^*)a),$$
      and since $\cll(\cla_{\infty,\tau}) \subseteq
      \cla_{\infty,\tau},$ the first two terms in the right hand
      side of  the above expression are weakly continuous in $a$,
      so we are left with the term
      $\tau(\delta(x^*)^*\delta(x^*)a)$. Let us choose an
      approximate identity   $e_n$ of the  $C^*$ algebra
      $\cla$ such that each $e_n$ belongs to $\cla_\tau$ (this is
      clearly possible, since $\cla_\tau$ is a norm-dense $\ast$-ideal, and for $z \in \cla_\tau$,
        one has that $|z| \in \cla_\tau$). By normality of $\tau$, $ \tau(\delta(x^*)^*\delta(x^*))
      =sup_n \tau(e_n\delta(x^*)^*\delta(x^*)e_n)=2sup_n
        \| \delta_0(x^*)e_n \|_\clk^2 \leq 2 \|e_n \|_{\infty, 0}^2
        \eta (x^*,x^*) < \infty,$ since $\| e_n \|_{\infty,0} \leq
        1$ and $x^* \in \cla_{\infty,\tau} \subseteq Dom(\eta)$.
        Thus, $\delta(x^*)^*\delta(x^*)=y^2$ for some $y \in
        \cla_\tau$, hence
        $\tau(\delta(x^*)^*\delta(x^*)a)=\tau(yay)$ with $y \in
        \cla_\tau$, which proves the required weak continuity.

  \qed

   Now we obtain the Christensen-Evans type form of the generator $\cll$.
\bthm
\label{ceform}
Let $R : h \raro \clk$ be defined as follows :
 $$ \cld(R)=\cla_{\infty,\tau},~~Rx \equiv \sqrt{2} \dlt_0(x).$$
 Then $R$ has a densely defined adjoint $R^*$, whose domain contains the linear span of
 the vectors $\dlt(x)y,~x,y \in \cla_{\infty,\tau}$ and
$$R^*(\dlt(x)y)= x\cll(y)-\cll(x)y-\cll(xy).$$
 We denote the closure of $R$ by the same notation $R$.  For $x,
  y \in \cla_{\infty,\tau}$,
$$ (R^* \pi(x) R -\frac{1}{2}R^*R x -\frac{1}{2}x R^*R)(y)=\cll(x)
y.$$
 Furthermore, $$\dlt(x)y=(Rx-\pi(x)R)(y), x,y \in
 \cla_{\infty,\tau},$$  $$\cll_2=-\frac{1}{2}R^*R.$$
 \ethm
 {\it Proof :-}\\ For $x,y,z \in \cla_{\infty,\tau},$
we observe by using the symmetry of $\cll$ that \bean \lefteqn{
\lgl \dlt(x)y, Rz \rgl}\\ &=& 2\lgl \dlt_0(x)y,\dlt_0(z)\\ &=&
\tau(y^*\cll(x^*z)-y^*\cll(x^*)z-y^*x^* \cll(z))\\ &=&
\tau(\cll(y^*)x^*z-(\cll(x)y)^*z-\cll(xy)^*z )\\ &=& \lgl \{
x\cll(y)-\cll(x)y-\cll(xy) \},z \rgl. \eean
 This suffices for the proof of the statements regarding $R^*$. It  can
 be verified by a straightforward computation that
 $(R^* \pi(x) R -\frac{1}{2}R^*R x -\frac{1}{2}x R^*R)(y)=\cll(x) y$ holds for
  $x ,y\in \cla_{\infty,\tau}$. The
  remaining statements are also verified in a straightforward
  manner.
\qed

\section{HP Dilation}
We shall now prove the existence of a unitary HP dilation for $T_t$.

  \bthm
  \label{hpcov}
  There exist a Hilbert space $k_1$ and a partial isometry $\Sigma : \clk \raro h \ot k_0$ (where
  $k_0=L^2(G)\ot k_1$) such that $\pi(x)=\Sigma^*(x \ot I_{k_0})\Sigma$ and  $\tilde{R} \equiv \Sigma  R $ is covariant in the sense
    that $(u_g \ot v_g) \tilde{R}= \tilde{R}u_g  $ on $\cla_{\infty,\tau}$    where $v_g=L_g \ot I_{k_1}$, $L_g $ denoting the left regular representation
    of $G$ in $L^2(G)$.
    \ethm
    Proof :\\
    The proof is essentially by the ideas as those in \cite{CGS}, so we omit the details.
     First we construct a strongly continuous unitary representation
          $V_g$ of $G$ in $\clk$ (strong continuity will follow by covariance of $\cll$ on a dense set
           of vectors, and hence by unitarity for every vector) such that  $\pi$
            is covariant under this $G$-action in $\clk$. This $V_g$ satisfies $V_g \dlt(x)=\dlt(\al_g(x))$ by the
             construction, which clearly implies that $V_g R=R u_g$ on $\cla_{\infty,\tau}$.
             Thus, $\pi$ is a normal covariant $\ast$-representation of $\bar{\cla}$ in $\clk$, hence extends to a normal
              $\ast$-representation, say $\bar{\pi}$ of the crossed product von Neumann algebra $\cla >\!\!\!\lhd G $, which is the weak closure
               of the algebra generated by $(x \ot I_{L^2(G)}), x \in \bar{\cla}$ and $u_g \ot L_g, g \in G$ in $\clb(h \ot L^2(G))$.
                Thus there is $\Sigma : \clk \raro h \ot L^2(G) \ot k_1$ (for some $k_1$) such that $\Sigma^*(X \ot I_{k_1})\Sigma
                =\bar{\pi}(X),$ for $X \in \bar{\cla}>\!\!\!\lhd G$. So in particular $\Sigma^*(x \ot I_{k_0})\Sigma=\pi(x)$, and
                 $\Sigma^*(u_g \ot v_g)\Sigma =V_g$. The rest of the proof follows easily from the arguments similar to those
                  in \cite{CGS}.

             \qed

         It is clear that for $x \in \cla_{\infty,\tau}$, $\cll(x)=\tilde{R}^*(x \ot 1_{k_0})\tilde{R}-
 \frac{1}{2}\tilde{R}^* \tilde{R} x-\frac{1}{2}x \tilde{R}^* \tilde{R}.$ This enables us to write down the candidate for
 the unitary dilation for the q.d.s. $T_t$.

  Before stating and proving the main theorem concerning H-P dilation, we make a crucial observtion. Let us consider
the form-generator given by $\clb(h) \ni x \mapsto \lgl {\tilde R}u, (x \ot 1){\tilde R}v \rgl -\frac{1}{2} \lgl xu, {\tilde R}^*{\tilde R} v \rgl -
\frac{1}{2} \lgl {\tilde R}^*{\tilde R}u, xv \rgl,~~ u, v \in \cld({\tilde R}^*{\tilde R}).$ By the construction of Davies (\cite{Dav}), there exists a unique
 minimal q.d.s. on $\clb(h)$, say $\tilde{T_t}$, such that the predual semigroup of $\tilde{T_t}$, say
 $\tilde{T_{t,*}}$,  has the generator (say $\tilde{\cll}_*$) whose domain  contains
 all elements of the form $y=(1+{\tilde R}^*{\tilde R})^{-1}\rho(1+{\tilde R}^*{\tilde R})^{-1}$
 for $\rho \in \clb_1(h)$, and $\tilde{\cll}_*(y)=
 \pi_*({\tilde R}_1 \rho {\tilde R}_1^*)-\frac{1}{2}{\tilde R}_1^*
 {\tilde R}_1 \rho-\frac{1}{2}
 \rho {\tilde R}_1^*{\tilde R}_1,$ where ${\tilde R}_1={\tilde R}(1+{\tilde R}^*
 {\tilde R})^{-1}$ and $\pi_*$ denotes the predual of the normal $\ast$-representation
  $x \mapsto (x \ot 1)$ of $\clb(h)$ into $\clb(h \ot k_0)$ (i.e. for $T \in \clb_1(h \ot k_0)$,
   $\pi_*(T)=\sum_i T_{ii}$, $T_{ii} \in \clb_1(h) $ being the diagonal elements of $T$ expressed
    in a block-operator form w.r.t. an o.n.b.of $k_0$, and the sum is in the trace-norm).
\blmma \label{conservative} $\tilde{T_t}$ is conservative. \elmma
{\it Proof :}\\ Let $\tilde{\cll}$ denote the generator of
$\tilde{T_t}$. We claim that $\cla_{\infty,\tau} \subseteq
\cld(\tilde{\cll})$ and $\tilde{\cll}=\cll$ on
$\cla_{\infty,\tau}$. Fix any $x \in {\cla}_{\infty,\tau}$.
  Let $\cld_*$ be the
 linear span of operators of the form $(1+{\tilde R}^*{\tilde R})^{-1}\sigma
  (1+{\tilde R}^*{\tilde R})^{-1}$ for $\sigma \in \clb_1(h)$. Clearly, for $\rho
 \in \cld_*$, $tr(\tilde{\cll}(x)\rho)=tr(x \tilde{\cll}_*(\rho))=tr(\cll(x)\rho)$ (using the explicit
  forms of $\cll$ and $\tilde{\cll}$), and since $\cld_*$ is a core for $\tilde{\cll}_*$ (see
 \cite{Dav}), we have   $tr(x \tilde{\cll}_*(\rho))=tr(\cll(x)\rho)$ for all $\rho \in \cld(\tilde{\cll}_*).$ Now, for
 $\rho \in \cld(\tilde{\cll}_*),$
$tr(\frac{\tilde{T_t}(x)-x}{t}
\rho)=tr(x(\frac{\tilde{T_{t,*}}(\rho)-\rho}{t})=tr(x
\tilde{\cll}_*(t^{-1} \itt_{0}^{t} \tilde{T_{s,*}}(\rho)ds))=
tr(\cll(x)t^{-1}\itt_{0}^{t}\tilde{T_{s,*}}(\rho)ds));$ and we
extend this equality by continuity to all $\rho \in \clb_1(h).$
Letting $t \raro 0+$, we get that $x \in \cld(\tilde{\cll})$ and
$tr(\tilde{\cll}(x)\rho)=tr(\cll(x)\rho) \forall \rho \in
\clb_1(h)$, which implies that $\tilde{\cll}(x)=\cll(x)$.
  From this, it follows by easy arguments using the fact that the resolvents of $\cll$ leaves
   $\cla_{\infty,\tau}$ invariant that $\tilde{T_t}(x)=T_t(x)~\forall x \in {\cla}_{\infty,\tau}$,
     and hence by the ultraweak density of
  $\tilde{\cla}_{\infty,\tau}$ in $\bar{\cla}$, $T_t$ and $\tilde{T_t}$ agree on $\bar{\cla}$ (where
   we use the same notation for the $C^*$ semigroup $T_t$ and its canonical normal extension on $\bar{\cla}$).
    In particular $\tilde{T_t}(1)=1.$
\qed

       We note that since the set of smooth complex-valued functions on $G$ with compact supports is dense in
        $L^2(G)$ in the $L^2$-norm, it is clear that ${k_0}_\infty$ is dense in the Hilbert space $k_0$, so let us
         choose and fix an orthonormal basis $\{e_i \}$ of $k_0$
          from $k_{0,\infty}$.  (note that $k_0$ can of course be chosen to be
          separable since $\bar{\cla}$ is $\sigma$-finite von Neumann algebra and $G$ is second countable )

         \bthm
         \label{hpdil}

          The q.s.d.e. $$dU_t=U_t(a^\dag_{\tilde{R}}(dt)-a_{\tilde{R}}(dt)-\frac{1}{2}{\tilde R}^*{\tilde R}dt);U_0=I$$
          on the space $h \ot \Gamma(L^2(R_+)\ot k_{0})$ admits a unique {\bf unitary} operator valued solution which implements a
           HP dilation for $T_t$.\\
           \ethm

                  For the meaning of such q.s.d.e. with unbounded coefficients we refer to \cite{MoS}
                   and \cite{Fag}  and for the notation $a_{\tilde{R}}
                   ,a^\dagger_{\tilde{R}}$ we refer to
                   \cite{GS}.\\

         Proof :-\\
         Since  ${\tilde R}^*{\tilde R}=-2\cll_2$, and since $\h \subseteq \cld(\cll_2)
          \subseteq \cld(\tilde{R})$, the closed Hilbert space operator $\tilde{R}$ is
           also continuous as a map from $\h$ to
           $h \ot k_0$ w.r.t the Frechet topology and the Hilbert space topology of the domain and the range
           respectively.
            Thus the relation ${\tilde R}u_g=(u_g \ot v_g){\tilde R}$ on $\cla_{\infty,\tau}$ extends by continuity to
             $\h$. That is,
             $\tilde{R}$ is covariant, and by the assumptions made on $\cll_2$ at the beginning
              of this section it is easy to see that the conditions of the Theorem \ref{*cont}
               are satisfied, so that  there are  $C,p$ such that
               $\|\tilde{R}\xi\|_{2,0} \leq C \| \xi \|_{2,p}.$
                Moreover,
                by \ref{*cont}, we obtain in particular that
                $Dom({\tilde R}^*)$(Hilbert space domain) contains $(h \ot k_0)_\infty$.
                 We recall from \cite{GS} the notation $\lgl f,S \rgl$ (where $S$ is a linear map from
                  $h$ to $h \ot k_0$ and $ f \in k_0$), which is defined to be a linear map
                   from $h$ to itself with the same domain as that of $S$, and satisfying $\lgl
                   \lgl f,S\rgl \beta,\gamma \rgl = \lgl S \beta, \gamma \ot f
                   \rgl$.
                       Now, for any vector
                 $f \in {k_0}_\infty$, it is clear that $\h \subseteq
                 Dom(<f,\tilde{R}>^*)$.
                   So in particular, we have that $\h \subseteq Dom({\tilde R}_i), Dom({\tilde R}_i^*)$,
                    where $\tilde {R}_i=<e_i,\tilde{ R}>$.  It is easy to note that $\tilde{R}_i,\tilde{R}_i^*$
                     keep $\h$ invariant.

                     We shall now use the  results of \cite{MoS}. Let $\clz_R$ denote the
                      class of elements $L\equiv (( L^i_j ))_{i,j \geq 0}, L^i_j \in \clb(h)$
                       such that for each $j$ there is $c_j$ with $\sum _i || L^i_j v \|_{2,0}^2
                        \leq c_j^2 \| v \|_{2,0}^2 \forall v \in h$. For $L \in \clz_R$, we define
                         $\cll^i_j \equiv L^i_j+(L^i_j)^*+\sum_{k \geq 1} (L^k_i)^*L^k_j$,
                          $\cli_R\equiv \{ L \in \clz_R : \cll^i_j=0 \forall i,j \}$, and $\clz_R^-
                          \equiv \{ L \in \clz_R : \cll_{S^\prime} \leq 0, S^\prime
                            \subseteq N, {\rm card}S < \infty \}$, where $\cll_{ S^\prime} \equiv
                             (( \cll^i_j ))_{i,j \in S^\prime}$. For any family $L=(( L^i_j ))_{i,j \geq 0}
                             $ where $L^i_j$ are closed densely defined operators, we define $\tilde{L}\equiv
                              (( \tilde{L}^i_j ))$, where $\tilde{L}^i_j=(L^j_i)^*$, and we set
                               $\tilde{\clz}_R= \{ L : \tilde{L} \in \clz_R \}$ and similarly define
                                $\tilde{\clz}_R^-,\tilde{\cli}_R$. Given a dense subspace $\cld$ of $h$,
                                 we denote by $\clz^-(\cld)$ the class of elements $Z\equiv ((
                                 Z^i_j ))$ such that $Z^0_0$ is the generator of a strongly continuous
                                  contractive semigroup on $h$ with $\cld$ as a core for $Z^0_0$,
                                   $\cld \subseteq Dom(Z^i_j) \forall i,j$, and there is a sequence
                                    $Z(n) \in \clz_R \bigcap \tilde{\clz}_R $ with $\lim_{n \raro \infty}
                                     Z^i_j(n) v =Z^i_j v \forall v \in \cld$. We define the bilinear forms
                                      $\cll^i_j(X)$ for $X \in \clb(h)$ by setting
                                       $$ \lgl \beta, \cll^i_j(X)\gamma \rgl =\lgl \beta,XZ^i_j \gamma \rgl
                                       +\lgl Z^j_i \beta,X \gamma \rgl +\sum_{k \geq 1} \lgl
                                        Z^k_i \beta, XZ^k_j \gamma \rgl.$$ We also set
                                         $\cli = \{ Z \in \clz^-(\cld): \cll^i_j(I)=0 \forall i,j \},
                                          \tilde{\cli}=\{ Z \in \clz^-(\cld) : \tilde{Z} \in \cli \},$
                                           and furthermore $\beta_\lambda = \{ X \geq 0, X \in \clb(h):
                                            \cll^0_0(X)=\lambda X \}$ for $\lambda>0$ ,
                                             and
                                             $\tilde{\beta}_\lambda$
                                             is defined by
                                             replacing $Z$ by
                                             $\tilde{Z}$. With
                                             these notations, we
                                             have  from \cite{MoS}      that  the
                                             q.s.d.e.
                                             $dU_t=\sum_{i,j}
                                             V_tZ^i_jd\Lambda^j_i(t);
                                             V_0=I$ has a unique
                                             unitary solution
                                             provided $Z \in \cli
                                             \bigcap \tilde{\cli}$
                                             and
                                             $\beta_\lambda=\tilde{\beta}_\lambda=\{0\}.$

                                                 We now
                                                      take
                      $\cld=\h$, $Z^i_j=0$ for $i,j \geq 1, $ $Z^i_0=\tilde{R}_i$,
                     $Z^0_i=\tilde{R}_i^*$ and $Z^0_0=\cll_2=-\frac{1}{2}{\tilde R}^* \tilde {R}$. Let $G_n=n(n-\cll_2)^{-1}$.
                      Let $Z^i_j(n)=G_nZ^i_j G_n$. We shall show that this choice satisfies
                      properties described above.
                        We first note that $G_n$ is clearly a bounded (with $\|G_n\|\leq 1$) covariant map, hence smooth
                        of order $0$ with bound $\leq 1$, in particular maps $\cld$ into itself. We have that
                         for $\xi \in \cld$,
                        $$
                         \sum_i \|Z^i_0(n)\xi\|_{2,0}^2 \leq \| (-2\cll_2)^{\frac{1}{2}} \xi\|^2_{2,0}
                         \\
                          $$
                          (as $R_i^*G_n^2R_i \leq R_i^*R_i$ and $G_n, \cll_2$  commute).

                           Similarly we find that $
                             \sum \|Z^i_0 \xi \|_{2,0}^2 <\infty$. For $j >0$, we also have that $ \| Z^0_j (n)\xi \|_{2,0}^2
                             =\| G_n \tilde{R}^*_jG_n \xi \|_{2,0}^2 \leq C^2_j \|\xi \|_{2,p_j}^2$, where we have used
                              Theorem \ref{*cont} to get constants $C_j,p_j$ such that
                         $\| \tilde{R_j}^* \xi \|_{2,0} \leq C_j \| \xi\|_{2,p_j} $
                         for $\xi \in \cld$.
                          Finally we verify that $\lim_{n \raro \infty}
                                     Z^i_j(n) v =Z^i_j v \forall v \in \cld$.  This follows from the following general fact :\\
                          If $L$ is a closed linear map from $h$ to $h$ with $\h$ in its domain, so that $\| L \xi\|_{2,0} \leq
                           M \| \xi \|_{2,r}$ for some $M,r$, then for $\xi \in \h$, $G_n L G_n \xi \raro L\xi$ as $n \raro \infty$.
                            To prove this fact, it suffices to observe that $G_n \xi$ clearly in $\h$ and  $\|G_n \xi-\xi\|_{2,r}^2
                            =\sum_{i_1,i_2,...i_k;k\leq r} \|(G_n-I)(d_{i_1}d_{i_2}...d_{i_k}\xi)\|_{2,0}^2$ (as $G_n$ is covariant),
                             which goes to $0$ as $G_n \raro I$ strongly. Thus we have $\|G_nLG_n \xi -L\xi\|_{2,0} \leq \|G_nL(G_n \xi -\xi)\|_{2,0}
                             +\|(G_n-I)L\xi \|_{2,0} \leq M\|G_n\xi -\xi \|_{2,r}+\| (G_n -I)L \xi \|_{2,0} $, which completes the proof of
                              the fact. Now, the existence, uniqueness and unitarity of $U_t$ follows from \cite{MoS} , noting the facts that
                               $\tilde{Z}=Z$ in this case and the conservativeness  of $\tilde{T_t}$ proved earlier suffices for
                                $\beta_\lambda =\tilde{\beta_\lambda}={0}$.
                                \qed

\newsection{Evans Hudson type Dilation}
We now study sufficient conditions for the existence of Evans-Hudson
(E-H)
dilation for the q.d.s $T_t$ considered
 in the previous section. We shall make the following additional assumptions (either
 {\bf A4},{\bf A5},
  {\bf A6},{\bf A7}; or {\bf A4},{\bf A5},{\bf A6},{ $\bf A7^{\prime}$})
 on the algebra and the group action:\\
 {\bf A4:}  $\cla_0=\{ x \in \cla_\infty : \exists C_x >0 s.t.
 \|x\|_{\infty,n}\leq \|x\|_{\infty,0}C_x^n ~\forall n \}$
 is \\norm-dense in $\cla$;\\
 {\bf A5:} $h_0=\{ u \in  h_\infty : \exists M_u >0 s.t.
 \|u\|_{2,n}\leq \|u\|_{2,0}M_u^n ~\forall n \}$ is $L^2$-dense in $h$;\\
 {\bf A6:} The canonical homomorphism of the crossed-product von Neumann algebra $\bar{\cla} >\!\!\!\lhd G$
    onto  the weak closure of the $\ast$-algebra generated by $\{ \bar{\cla},u_g;g\in G\}$
     in $\clb(h)$ is an isomorphism;\\
     and
 either \\
 {\bf A7:} Assume that the trace $\tau$ is finite
(hence $\cla_\infty$
 coincides with $\cla_{\infty,\tau}$) and we require the following. \\
  Let $\bar{\del_i}$ denote the  generator of the one-parameter group
 $\bar{\cla} \ni x \mapsto u_{g_t}x u_{g_t}^*$, w.r.t. the operator norm topology of $\bar{\cla}$,
  where $g_t=exp(t\chi_i)$, i.e. $x \in \bar{\cla}$
  belongs to the domain of $\bar{\del_i}$ if and only if $\frac{u_{g_t}xu_{g_{t}}^*-x}{t} $ converges
   as $t \raro 0+$ in the operator norm topology of $\bar{\cla}$ inherited from $\clb(h)$.
    Let $\bar{\cla}_\infty$ denote the   intersection of the domains of
     $\bar{\del_{i_1}}... \bar{\del_{i_k}}$ for all possible choices of $i_1,...i_k$.
     We assume that $\bar{\cla}_\infty=\cla_\infty$, i.e. any element of $\bar{\cla}$ which
      belongs to the intersection of all the domains (``smooth"), then it must belong to
      $\cla$.\\
     or\\
        { $\bf A7^\prime$:} The trace $\tau$ is not finite, and in this case we assume that
         $\{ \bar{\cla},u_g;g \in G \}^{\prime \prime}$ is a type I
         factor
         isomorphic with $\clb(h)$.

    We  note that {\bf A4} and {\bf A5} hold for $G$ compact or $G=R^n$.
     For compact groups, the
                                     verification of {\bf A4},{\bf A5} can be
                                     done by the Peter-Weyl
                                     decomposition, and for
                                     $R^n$, the space
                                     $\cls(R^n)$ of Schwarz
                                     functions can be taken as a
                                     candidate for both $\cla_0$
                                     as well as $h_0$. Similar
                                     thing will be true for many
                                     more interesting noncompact
                                      Lie groups.
 Furthermore,
    {\bf A6} implies that the similar statement holds with $\bar{\cla}$ replaced
     by $\cla^\prime$.

     \blmma
     \label{crossedproduct}
     Under the assumption ({\bf A6}), it follows that $\cla^\prime >\!\!\!\lhd G$ is isomorphic with
       $\{ \cla^\prime,u_g;g\in G\}^{\prime \prime}$.
      \elmma
      {\it Proof:-}\\
      The proof is an easy consequence of the Tomita-Takesaki modular theory. Let $\clj$
       denote the closed extension of the canonical anti-linear isometric involution of that theory which sends
        $x \in \cla_{\infty,\tau}$ to $x^*$. Since $u_g(x)=\al_g(x)$ for $x \in \cla_{\infty,\tau}$,
         and $\al_g$ is $\ast$-preserving, it is clear that $u_g\clj=\clj u_g$ and thus (since we also
          have $\clj^2=id$,) $\clj u_g \clj=u_g$. But we know from the Tomita-Takesaki theory that
           $\clj \bar{\cla} \clj=\cla^\prime$, hence the von Neumann algebra $\clb\equiv\{\bar{\cla},u_g;g\in G \}^{\prime
           \prime}$ is anti-isomorphic under the map $\clj . \clj$ with $\clc \equiv
            \{ \cla^\prime,u_g;g\in G
            \}^{\prime \prime}$. But by assumption {\bf A6}, $\clb$ is isomorphic with $\bar{\cla}
            >\!\!\!\lhd G$, which is nothing but the von Neumann algebra generated by $\bar{\cla} \ot 1$
             and $u_g \ot L_g$ in $\clb(h \ot L^2(G))$, where $L_g$ is the left regular representation.
              Clearly $\bar{\cla} >\!\!\!\lhd G$ is anti-isomorphic with $\cla^\prime >\!\!\!\lhd G$, under the obvious
               anti-isomorphism $(\clj . \clj)\ot id$. Thus $\cla^\prime >\!\!\!\lhd G$ is isomorphic
                with $\clc$, since the composition of two anti-isomorphism is an isomorphism.
                \qed

   \bthm
   \label{ehstructure}
   Under the assumption  {\bf A6} above, there exist a Hilbert space $k_0$, a partial isometry
    $\tilde{\Sigma} : h \ot k_0 \raro h \ot k_0$ and a closed linear map  $\tilde{S}$ from
     $h$ into $h \ot k_0$ with $h_\infty$ in its domain, such that the followings hold :\\
     (i) $\tilde{\Sigma}$ is covariant, i.e.  $\tilde{\Sigma}(u_g \ot id)=(u_g \ot id)\tilde{\Sigma}$;\\
      (ii) $\tilde{\pi}(x)\equiv\tilde{\Sigma}(x \ot 1_{k_0})\tilde{\Sigma}^*$ is a covariant normal
       $\ast$-homomorphism of $\bar{\cla}$ into $\clb(h \ot k_0)$, which is structural
        in the sense of \cite{GS}, i.e. $\tilde{\pi}(\bar{\cla})\subseteq \bar{\cla}\ot \clb(k_0)$ and we also have
         $\tilde{\pi}(\al_g(x)) (u_g \ot id)=(u_g \ot id) \tilde{\pi}(x)$;\\
             (iii) $\tilde{S}$ from $h$ to $h \ot k_0$ is smooth covariant map (w.r.t.
                   the same $G$-action as in (i)),i.e. $\tilde{S}u_g=(u_g \ot id) \tilde{S}$; and $\tilde{\delta}(x)\equiv \tilde{S}x-\tilde{\pi}(x)
      \tilde{S}$ for $x \in \cla_{\infty,\tau}$ extends as a bounded map from $h$ to $h\ot k_0$
       which is also structural, i.e. $\tilde{\delta}(x) \in \bar{\cla} \ot k_0$, $x \in \cla_{\infty,
        \tau}$, and covariant in the sense that $\tilde{\delta}(\al_g(x))=\gamma_g(\tilde{\delta}(x))$,
         where $\gamma_g : \bar{\cla} \ot k_0 \raro \bar{\cla}\ot k_0$ by $\gamma_g(.)=(u_g \ot id).(u_g^* \ot id).$;\\
      (iv) $\cll(x)=\tilde{S}^* \tilde{\pi}(x)\tilde{S}-\frac{1}{2}\tilde{S}^*\tilde{S}x
      -\frac{1}{2}x \tilde{S}^*\tilde{S}$ for $x \in \cla_{\infty,\tau}$, in the sense that the LHS
        is a bounded extension of the RHS, which is defined on its natural domain.\\
 (v) Furthermore, if we also make the assumption {\bf A7}, then we have a stronger structurality
  in the following sense :\\
  For $x \in \cla_\infty$ and $\xi,\eta \in k_0$, $<\xi, \tilde{\delta}(x)>\in \cla_\infty$ and
   $<\xi, \tilde{\pi}(x)\eta>\equiv <. \ot \xi,\tilde{\pi}(x)(. \ot \eta)> \in \cla_\infty$.
      \ethm
      {\it Proof :-}\\
       First of all we proceed in the line of the proof of Theorem  \ref{hpcov}, and consider
        the covariant normal  $\ast$-homomorphism $\pi$ as in that theorem and lift it to a normal $\ast$-homomorphism
         $\bar{\pi}$ of the crossed product, which is isomorphic by the assumption with $\{
          \bar{\cla},u_g;g\in G \}^{\prime \prime}$, and thus we take $\bar{\pi}$ to be a normal $\ast$-homomorphism
           on       $\{
          \bar{\cla},u_g;g\in G \}^{\prime \prime}$ satisfying $\bar{\pi}(x)=\pi(x)$ for $x\in \bar{\cla}$
           and $\bar{\pi}
           (u_g)=V_g$. Then the construction  as in \ref{hpcov} ensures that there exist
            some Hilbert space $k_1$ and    isometry $\Sigma$ from $\clk$ to $h \ot k_1$
             (where $\clk$ is as in the proof of \ref{hpcov}) such that $\Sigma^*(x\ot 1_{k_1})\Sigma
              =\bar{\pi}(x), \Sigma^*(u_g \ot 1_{k_1})\Sigma=V_g$. Let $\clk_1 \equiv \Sigma(\clk)
               \subseteq h \ot k_1$, and let $P_1=\Sigma \Sigma^*$ and $\delta_1(x)=\Sigma \delta(x),
                x \in \cla_{\infty,\tau}$. Clearly $\{ \delta_1(x)v,x\in \cla_{\infty,\tau},v \in h
                \}$ is dense in $\clk_1$. As in \cite{CGS},  we now construct
                 a normal $\ast$-homomorphism $\pi^\prime$ of $\cla^\prime$ in $\clb(\clk_1)$ by setting
                  $\pi^\prime(a)\delta_1(x)v=\delta_1(x)av$, and extending by linearity and continuity (details
                   can be found in \cite{CGS} or \cite{GS}). We extend this $\pi^\prime(a)$ on the whole of
                    $h\ot k_1$ by putting it equal to $0$ on $P_1^\perp(h \ot k_1)$, and we denote this
                     trivial extension also by $\pi^\prime(a)$. Clearly, $\pi^\prime$ is covariant w.r.t. $g \mapsto
                   (u_g \ot 1)$, and thus we can extend $\pi^\prime$ to a normal $\ast$-homomorphism of
                    $\{ \cla^\prime, u_g; g\in G \}^{\prime \prime}$ (which is isomorphic with the crossed
                     product von Neumann algebra $\cla^\prime >\!\!\!\lhd G$), say $\pi^{\prime \prime}$, satisfying
                      $\pi^{\prime \prime}(u_g)=u_g \ot 1_{k_1}$. Hence there exist a Hilbert space $k_2$
                       and a partial isometry $\Sigma_1 : h \ot k_1 \raro h \ot k_2$ such that $\Sigma_1^*
                        (a \ot 1_{k_2})\Sigma_1=P_1(a \ot 1_{k_1})$ and $\Sigma_1^*(u_g \ot 1_{k_2})
                         \Sigma_1=P_1(u_g \ot 1_{k_1})$.  We take $k_0=k_1 \oplus k_2$, $\tilde{\Sigma}
                         =\Sigma_1 \oplus 0|_{h\ot k_2}: h \ot k_0 \equiv (h\ot k_1) \oplus
                          (h \ot k_2) \raro h \ot k_0$, and $\tilde{S}v:=\tilde{\Sigma}(\Sigma Rv \oplus 0)$.
                           The remaining is verified as in \cite{CGS}, and hence the details are omitted.

                                  Finally, to prove (v), we first note that since $\tilde{\Sigma}$ is covariant and bounded
                                   (in Hilbert space sense), $u_{g_t}<\xi,\tilde{\pi(x)}\eta>u_{g_t}^*=<\xi,\tilde{\pi}(\al_{g_t}(x))\eta>$,
                                     (where $g_t=exp(t \chi_i)$, for some fixed $i$ ) and thus we have that for $x \in \cla_\infty$,
                                     $\frac{u_{g_t}<\xi,\tilde{\pi}(x)\eta>u_{g_t}^*-<\xi,\tilde{\pi(x)}\eta>}{t}
                                     -<\xi,\tilde{\pi} ( \del_i(x))\eta>=t^{-1}\int_0^t <\xi, \tilde{\pi}(\al_
                                     {g_s}(\del_i (x))-\del_i(x))\eta > ds;
                                      $ which goes to $0$ in norm as
                                     $\tilde{\pi}$ is a norm-contractive map.
                                       This shows that $<\xi, \tilde{\pi}(x)\eta>$
                                     belongs to the domain of $\bar{\del_i}$ for any $i$, and repetition of similar arguments
                                      proves that it belongs to $\bar{\cla}_\infty=\cla_\infty$. We can prove similar fact
                                       about $\tilde{\delta}$ by using the covariance of $\tilde{\delta}$.
                                       We first note that $\cll$,
                                       being covariant,
                                       norm-closed and having
                                       $\cla_\infty$ in its
                                       domain, is continuous in
                                       the Frechet topology. Using
                                       the Frechet continuity of
                                       $\cll$ and the cocycle
                                       identity
                                       $\tilde{\dlt}(x)^*\tilde{\dlt}(x)=\cll(x^*x)-\cll(x)^*x-x^*\cll(x)$ for $x \in \cla_\infty$,
                                        (by Assumption {\bf A7} $\cla_{\infty,\tau}=\cla_\infty$) we conclude that
                                         $\tilde{\dlt} :
                                         \cla_\infty \raro
                                         \clb(h,h \ot k_0)$ is
                                         continuous w.r.t. the
                                         Frechet topology on the
                                         domain and the operator
                                         norm topology on the
                                         range. Thus, if we denote
                                         by $\Psi$ the map
                                         $\cla_\infty \ni x \raro
                                         \lgl \xi, \tilde{\dlt}(x)
                                         \rgl$ (for fixed $\xi \in
                                         k_0$), then $\Psi :
                                         \cla_\infty \raro
                                         \clb(h)$ is continuous
                                         w.r.t. the Frechet
                                         topology of $\cla_\infty$
                                         and the norm topology of
                                         $\clb(h)$. Since
                                         $\frac{\al_{g_t}(x)-x}{t}
                                         \raro \partial_i(x)$ (as
                                         $t \raro 0+$) in the
                                         Frechet topology for $x
                                         \in \cla_\infty$, it
                                         follows that
                                         $\frac{u_{g_t} \Psi(x)
                                         u_{g_t}^*-\Psi(x)}{t}
                                         =\Psi(\frac{\al_{g_t}(x)-x}{t})
                                         \raro
                                         \Psi(\partial_i(x))$ as
                                         $t \raro 0+$ in norm,
                                         i.e. $\Psi(x) \in
                                         Dom(\bar{\partial_i})$
                                         with
                                         $\bar{\partial_i}(\Psi(x))=\Psi(\partial_i(x)).$
                                         Repeated use of such
                                         arguments proves that
                                         $\Psi(x) \in
                                         \bar{\cla}_\infty
                                         =\cla_\infty$, which
                                         completes the proof.
                                         \qed

                      We make the assumptions {\bf A1}-{\bf A7}(or {\bf A1}-{ $\bf A7^\prime$}) from now on for the rest of the section and
                          proceed to prove the existence of an Evans-Hudson dilation.
                       Our first step is to introduce a map $\Theta$, as in \cite{GS}, which combines
                        all the ``structural" maps $\cll, \tilde{\delta}, \tilde{\pi}$ into a single
                         map.  Let $\hat{k_0}\equiv C \oplus k_0$ and $\Theta :\cla_{\infty,\tau}
                          \raro \clb( h \ot \hat
                         {k_0}) $ given by $$  \Theta(x)=\left( \begin{array}{cc}\cll(x),\tilde{\delta}^\dagger
                         (x) \\ \tilde{\delta}(x),\tilde{\sigma}(x)\end{array} \right);$$
            where $\tilde{\delta}^\dagger (x)=\tilde{\delta}(x^*)^*$ and $\tilde{\sigma}(x)=
            \tilde{\pi}(x)-(x \ot 1_{k_0})$.
             \blmma
             \label{theta}
             There exist a Hilbert space $k_0^\prime$ and two covariant smooth maps $B :
              h \ot \hat{k_0} \raro h \ot k_0^\prime, $ $A: h \ot k_0^\prime \raro h \ot
               \hat{k_0}$ such that $\Theta(x)=A(x \ot 1_{k_0^\prime})B$, where covariance
                 is with respect  to  the representation $g \mapsto u_g \ot 1$ in all cases.
                \elmma
                {\it Proof :-}\\
                   We take $k_0^\prime=\hat{k_0}\ot C^3 \equiv \hat{k_0} \oplus \hat{k_0}
                    \oplus \hat{k_0}$. Let $A=(A_1, A_2,I), B=\left(\begin{array}{c}
                     B_1 \\ I\\ B_2 \end{array} \right)$, where $A_i,B_j$ s are covariant smooth maps
                      from $h \ot \hat{k_0}$ to itself given by (w.r.t. the direct sum decomposition
                       $h \ot \hat{k_0} = h \oplus (h \ot k_0)$):\\
                      $A_1=\left(  \begin{array}{cc} 0 & \tilde{S}^*\tilde{\Sigma}\\
                       0 & -\tilde{\Sigma} \end{array}\right) $, $A_2= \left( \begin{array}{cc}
                        -\frac{1}{2}\tilde{S}^*\tilde{S} & 0\\ \tilde{S} & -\frac{1}{2}1_{h \ot k_0}
                         \end{array} \right),$ $B_1=A_1^*$, $B_2=A_2^*$. We note that all the maps above
                          are defined with their usual domains and from the results of the previous
                           sections (since $\tilde{\Sigma}$ is a bounded covariant map and $\tilde{S}$
                            is smooth covariant, and satisfies the condition of \ref{*cont}, so
                             that its adjoint is smooth covariant too, and furthermore composition of
                              smooth covariant maps is again smooth covariant) it follows that $A$ and $B$
                               are indeed smooth covariant. That $\Theta(x)=A(x \ot 1_{k_0^\prime})B$ can
                                be verified by direct and easy computation.
                                \qed

               We now extend the definition of the map $\Theta$, taking advantage of the fact that
                $(A \ot I)$ and $(B \ot I)$ are also smooth covariant maps with the same order and bounds
                 as $A$ and $B$ respectively where $I$ denotes identity on any separable Hilbert space with trivial
                  $G$-action (see Lemma \ref{tensor} ).
                  \bdfn
                  \label{thext}
                  For any two Hilbert spaces $\clh_1,\clh_2$ and a smooth (not necessarily covariant)
                   map $X$ from the Sobolev-Hilbert space $(h \ot \clh_1, u_g \ot 1)$
                     to $(h \ot \clh_2, u_g \ot 1)$
                    we define $\Theta(X)$ to be a smooth map from $(h \ot \hat{k_0} \ot \clh_1, u_g \ot 1 \ot 1)$
                     to $ (h \ot \hat{k_0} \ot \clh_2, u_g \ot 1 \ot 1)$, given by,\\
                     $$ \Theta(X)=(A \ot 1_{\clh_2})P^\prime_{23}(X \ot 1_{k_0^\prime})P_{23}(B \ot 1_{\clh_1}), $$
                      where $P_{23}: h \ot k_0^\prime \ot \clh_1 \raro h \ot \clh_1 \ot k_0^\prime$ denotes
                      the operator which interchanges 2nd and 3rd tensor components, and a similar definition
                       is given for $P_{23}^\prime$.
                       \edfn

            It is clear that for $\clh_1=\clh_2=C$, we indeed recover the definition of $\Theta(x)$ for
             $x \in \cla_{\infty,\tau}$, and furthermore we make sense of the same symbol even for
              $x$ which are not necessarily bounded operator on $h$, but are smooth map on $h$ w.r.t.
               the $G$-action given by $u_g$. The above extended definition enables us to compose
                $\Theta$ with itself, i.e. we can make sense of $\Theta(\Theta(X))$ and so on. We
                 shall denote $\Theta(...\Theta(X) )$ ($n$-times composition) by $\Theta^n(X)$ for
                  $X$ as in the definition \ref{thext}. The following estimate will be  a fundamental tool
                   for proving the homomorphism property of the E-H dilation which we are going to construct.\\

      \bthm
      \label{thest}
      For $x \in \cla_\infty$, $\xi \in  (h \ot \hat{k_0}^n)_\infty$, (where $\hat{k_0}^n$
       denotes $n$-fold tensor product of $\hat{k_0}$ with itself, and $G$-action is taken
        to be $u_g \ot 1_{\hat{k_0}^n}$ ) we have that
        \be
        \label{basicest}
         \| \Theta^n(x)\xi \|_{2,m} \leq C_1 C_2^n \| x \|_{\infty,np_1+m}\| \xi \|_{2,n(p_1+p_2)+m};
         \ee
          where $N$ is the dimension of the Lie algebra of $G$, $C_1=\sqrt{\frac{2}{4N-1}}(2 \sqrt{N})^{m+1},$ $C_2=c_1c_2(2 \sqrt{N})^{p_1}$,
           $c_1,c_2,p_1,p_2$ are such that $A$ is a smooth map of order $p_1$ and bound $\leq c_1,
           $ and $B$ is a smooth map of order $p_2$ and bound $\leq c_2$. In particular, let $x \in \cla_0$,
              $\xi_n=v \ot w_1 \ot ... \ot w_n$ with $v \in h_0$, $w_i \in \hat{k_0}$,
              $\| w_i \|_{2,0} \leq K$ for some $K$, and let $V$ be a smooth covariant map of order $0$
               and bound $\leq 1$ on $(h \ot \hat{k_0}^n \ot \clk^\prime)_\infty $ (where $\clk^\prime$
                is some separable Hilbert space, with trivial $G$-action) and $\eta \in \clk^\prime$.
                  Then we have that  $\| (\Theta^n(x )\ot 1)V(\xi_n \ot \eta) \|_{2,0}\leq
               \| \eta \|_{2,0}K_1 K_2^n$ for some constants $K_2,K_2$ depending on $x,v,K$ only.

               Furthermore, for $x,y \in \cla_0$ and $v,V,\eta$ as above, we get constants $K_1^\prime,
                K_2^\prime$ depending on $x,y,v$ only such that
                $\| ((\Theta^n(x )^*\Theta^n(y))\ot 1)V(v \ot \eta) \|_{2,0}\leq
               \| \eta \|_{2,0}K^\prime_1 {K^\prime_2}^n$
      \ethm
{\it Proof :-}\\
First we note that for $x \in \cla_\infty$ and $v \in h_\infty,$ we have that
\be
\label{esteq}
 \| xv\|_{2,m} \leq \sqrt{\frac{2}{4N-1}}(2\sqrt{N})^{m+1}\|x\|_{\infty,m}\| v
 \|_{2,m}.
 \ee
  To prove this estimate  we note that for any fixed $k$-tuple $i_1,...i_k$ with $k \leq m$, where
   each $i_j\in \{ 1,...N \}$, we have that $ \| d_{i_1}...d_{i_k}(xv)\|_{2,0}^2=
    \| \sum_{J \subseteq \{1,...k \}} \del_J(x)d_{J^c}(v)\|_{2,0}^2$, where for any subset $J$
     of $\{ 1,...k \}$, we denote by $\del_J$ the map $\del_{i_{p_1}}...\del_{i_{p_t}}$,
      with $p_1<p_2<...p_t$ being the arrangement of the elements of $J$ ($|J|=t$) in the
       increasing order; and a similar definition is given for $d_{J^c}$, $J^c$ being the
        complement of $J$. Clearly, for any $l$ nonnegative numbers $a_1,...a_l$, one has that
         $ (a_1+...a_l)^2 \leq 2l (a_1^2+...a_l^2) $ and using this we see that
            $ \| \sum_{J \subseteq \{1,...k \}} \del_J(x)d_{J^c}(v)\|_{2,0}^2 \leq
             2^{k+1} \sum_J \| \del_J(x)d_{J^c}(v)\|_{2,0}^2 \leq 2^{k+1}
             \sum_J \| x\|_{\infty,m}^2
              \| v \|_{2,m}^2=2^{2k+1} \|x\|^2_{\infty,m} \|v \|_{2,m}^2  $. The proof of the estimate
                then follows by noting that the number of possible $k$-tuples as above is $N^k$, and
                $\sum_{k=0}^m 2^{2k+1} N^k=\frac{2}{4N-1}((4N)^{m+1}-1)$.

                Now we come to the proof of the main estimate in the present theorem.
           It is easy to see that the estimate (\ref{esteq}) holds even when $x \in \cla_\infty$
            is replaced by $(x \ot 1_\clh)$ for any separable Hilbert space $\clh$ and
             $v \in h_\infty$
             is replaced by $v \in (h \ot \clh)_\infty$ where
              $\clh$ carries  the trivial $G$-representation. Hence $\| \Theta^n (x) \xi \|_{2,m}
               \leq c_1^n \| (x \ot 1)P_{23}(B \ot 1)...P_{23}(B \ot 1)\xi \|_{2,m+np_1}
                \leq c_1^n \sqrt{\frac{2}{4N-1}}(2\sqrt{N})^{m+np_1+1}\|x\|_{\infty,m+np_1}
                 \|(B\ot 1)...P_{23}(B \ot 1)\xi\|_{2,m+np_1} $, from which the desired
                  estimate follows. (Note that $P_{23}$ and $1$ are used here as generic notation,
                    without distinguishing among different spaces on which they act, and also we have
                     made use of the fact that $P_{23}, P^\prime_{23}$ are trivially smooth covariant maps of order
                      $0$ and bound $\leq 1$.)

  The last two assertions of the result follows by noting that for $x \in \cla_0$, $v \in h_0$,
   we have that $\| x \|_{\infty, np_1+m} \leq \| x \|_{\infty,0}M_x^{np_1+m}$ and a similar
    estimate is valid for $v$, and furthermore we have that $\|V( \xi_n \ot \eta)\|_{2,n(p_1+p_2)+m}
     \leq \|v \|_{2,n(p_1+p_2)+m}K^n \| \eta \|^2$.
\qed

 We now prove an important algebraic property of $\Theta$.\\
 \blmma
 \label{thetaalg}
 For $x,y$ of the form $<\xi,\Theta^n(a)\eta>$ for some $a\in \cla_\infty, \xi, \eta \in
  (h \ot \hat{k_0}^{\ot^n})_\infty$, we have that
  $$ \Theta(xy)=\Theta(x)(y \ot 1)+(x\ot 1)\Theta(y)+\Theta(x)Q \Theta(y),   $$
   where $Q=\left( \begin{array}{cc} 0,0 \\0,1_{h \ot k_0} \end{array}
   \right)$as in \cite{GS}.
   \elmma
   {\it Proof :-}\\
    We extend the definition of $\tilde{\pi}$
    in the obvious manner to make sense of $\tilde{\pi}(x)$ for any $x \in \clb(h
    _\infty,h_\infty)$ so that $\tilde{\pi}(x)$ is also a smooth map. However, this
     extended $\tilde{\pi}$ need not be a homomorphism on whole of the $\clb(\h,\h)$.
      For proving the lemma, it suffices to verify that  $\tilde{\pi}(xy)
   =\tilde{\pi}(x)\tilde{\pi}(y)$ for $x,y$ of the form $<\xi,\Theta^n(a)\eta>$ for some $a\in \cla_\infty, \xi, \eta \in
  (h \ot \hat{k_0}^{\ot^n})_\infty$. We now consider the cases corresponding to the assumptions
   {\bf A7} and { $\bf A7^\prime$} separately.
    In the first case, we have already shown that $<\xi,\Theta^n(a)\eta>\in \cla_\infty$ for
     $a,\xi,\eta$ as in the statement of the lemma, and then it is trivial to see that the
      above algebraic relation holds, since $\tilde{\pi}$ is indeed a homomorphism on $\cla_\infty$.
       Now, we consider the case when { $\bf A7^\prime$}
        is assumed. In this situation,   the proof of the
        theorem \ref{ehstructure} enables us to see that $k_1,k_2,k_0$ (in the notation of
         the theorem mentioned above) can be chosen in such a way that $P_1=I_{h \ot k_1}$,
          and hence $\tilde{\Sigma}^*\tilde{\Sigma}=I_{h \ot k_1} \oplus 0_{h \ot k_2}$. But
           then it follows that $\clb(h_\infty,h_\infty) \ni x \mapsto
            \tilde{\pi}(x)=\tilde{\Sigma}(x \ot I_{h \ot k_1} \ot I_{h \ot k_2})\tilde{\Sigma}^*$ is a homomorphism.
             This completes the proof of the required algebraic property of $\Theta$ in the present
              case.
              \qed

\bthm
\label{impl}
Let $\tilde{\Sigma}, \tilde{S}, k_0$ be as in the statement of theorem \ref{ehstructure}. Then the
 following operator q.s.d.e. in $h \ot \Gamma(L^2(R_+,k_0))$ admits a contraction-valued
  solution.\\
  \be
  \label{ehimpl}
   dV_t=V_t \left( a_{\tilde{S}}^\dagger (dt)+\Lambda_{\tilde{\Sigma}-I}(dt)-a_{\tilde{\Sigma}^*
   \tilde{S}}(dt)-\frac{1}{2}{\tilde{S}^*}\tilde{S}(dt) \right); \\
   V_0=I.
   \ee
   Furthermore, $V_t$ is covariant w.r.t. the $G$-action $u_g \ot 1$ on $h \ot \Gamma(L^2(R_+,k_0))$.
\ethm
{\it Proof :-}\\
 The existence and uniqueness of the solution $V_t$ can be
 obtained essentially as in the Theorem \ref{hpdil} (see also
 \cite{MoS} and \cite{Fag}). Covariance of $V_t$ is
 straightforward to show. It is important to note that $V_t$ is
 not unitary since $\tilde{\Sigma}$ is a strict partial isometry.
     \qed

Now, let us recall the map-valued quantum stochastic calculus developed in \cite{GS}.
We want to define integrals of the form $\int_0^t Y_s \circ M(ds)$ where $M$
 stands for $(a_{\tilde{\delta}}(ds)+a^\dagger_{\tilde{\delta}}(ds)+\Lambda_{\tilde{\sigma}}(ds)+
 \cli_{\cll}(ds))$ (see notations in \cite{GS}), where $\tilde{\sigma}(x)=\tilde{\pi}(x)-(x \ot 1_{k_0})$.
  The ideas and construction will be almost the same as those in \cite{GS}, and hence we only briefly
   sketch the steps, omitting details. We shall adopt the following convention throughout
    the rest of the paper : unless otherwise stated, $G$-action on a Hilbert space of the form
     $h \ot \clh^\prime$ will be taken to be $u_g \ot 1$.

   In \cite{GS}, the integrator $\Theta$ was a norm-bounded map, and thus the above integral could
    have been defined on the whole of $\cla \ot \Gamma$ ($\Gamma =\Gamma(L^2(R_+,k_0))$).
     But here we are dealing with unbounded $\Theta$,
     and thus we shall make sense of the above
      integral only on a restricted domain. Let  $\cld$ denote the algebraic
       linear span of the elements of the form $x \ot e(f)$,
       with $x \in \clb(h_\infty,h_\infty)$,
        i.e. $x$ is a smooth map
        from $h_\infty$ to itself,
        $f$ is a bounded continuous $k_0$-valued function on $[0,\infty)$, and let $\cls$
         denote the space $\clb(h_\infty, (h \ot \Gamma)_\infty)$. Let $(Y_t)_{t \geq 0}$
          be a family of maps from $\cld$ to $\cls$, with the adaptedness in the obvious
           sense as in \cite{GS}, and also satisfying the following condition (an analogue of
            3.12 of \cite{GS}):\\
            \be
            \label{reg}
            \sup_{0 \leq s \leq t} \| Y_s(x \ot e(f))v\|_{2,0} \leq \|r_1 (x \ot 1_{\clh^\prime})r_2v\|_{2,0}\\
            \ee
        for some Hilbert space $\clh^\prime$, where $r_1,r_2$ are smooth maps between appropriate
         spaces, in contrast to their being bounded in \cite{GS}. We call such a process
          $Y_t$ regular, and we define $Z_t=\int_0^t Y_s \circ
          M(ds)$ exactly in the same way as in \cite{GS}. Indeed, $\Theta_{f(s)}(x)$ (with the already given
           definition of $\Theta$, and notations as in \cite{GS}) is a densely defined closable
            operator on $h$, with $h_\infty$ in its domain, and also $<f(s), \Theta(x){g(s)}>$ clearly
             is an element of $\clb(h_\infty,h_\infty)$ (i.e. smooth map), hence the conditions
              required in \cite{GS} for defining the maps $S(s),T(s)$ by $S(s)(ve(f_s))=\tilde{Y_s}(\tilde
              {\delta(x)}\ot e(f_s))v$ and $T(s)(ve(g_s)\ot f(s))=\tilde{Y_s}(\tilde{\sigma}(x)_{f(s)}
              \ot e(g_s))v$ as in \cite{GS} (with similar notation) are satisfied. Then
                  an analogue of Prop 3.3.5 of \cite{GS} can be applied to $Z_t$ (see
                  Cor. 2.2.4 (ii) of \cite{GS}),
                    and  we conclude
                   that $Z_t$ is well-defined and regular.  However, in the present
                    situation $Z_t(x \ot e(f))$ need not be a bounded operator from $h$ to $h \ot \Gamma$, and
                      need not belong to  $\bar{\cla} \ot \Gamma$.
                       Nevertheless, at least in case {\bf A7} is assumed, we shall  show that the EH flow $J_t$ to be constructed by us will
                         map $\bar{\cla} \ot \Gamma$ to itself.

                        \bthm
                        \label{ehmain}
                        We set $j_t : \bar{\cla}  \raro \clb(h  \ot \Gamma)$ by
                         $j_t(x) = V_t (x \ot 1_\Gamma) V_t^*$ (where $V_t$ as in theorem
                          \ref{impl} ) and also we define $J_t : \bar{\cla} \ot_{alg} \Gamma \raro
                           \clb(h,h \ot \Gamma)$ by $J_t(x \ot e(f))v=j_t(x)(ve(f))$ and extending by linearity.
                            Furthermore,  the above definition of  $j_t(x)$ can be extended to  $x \in \clb(h_\infty,\h)$
                            (i.e. for $x$ which are possibly unbounded as Hilbert space map)
                              and $J_t : \cld \raro \cls$ also similarly.
                            Then we have the following :\\
                            (i) $J_t$ is a regular process and satisfies the q.s.d.e. $dJ_t=J_t \circ M (dt);J_0=id$;\\
                            (ii) $j_t: \bar{\cla} \raro \clb(h \ot \Gamma)$ is a normal covariant (w.r.t. $\al_g \ot id$)
                              $\ast$-homomorphism which
                             dilates $T_t$ in the sense of \cite{CGS}, \cite{GS};\\
                             and\\
                             (iii) If {\bf A7} (and not { $\bf A7^\prime$}) is assumed,  then $j_t(\bar{\cla})\subseteq \bar{\cla}\ot \clb(\Gamma)$,
                             and\\
                              $<e(f),j_t(x){e(f^\prime)}> \in \cla_\infty$ for $x \in \cla_\infty, f,f^\prime \in
                               L^2(R_+,k_0)$.
                             \ethm
                             {\it Proof :-}\\
                             Since $V_t$ and $V_t^*$ are smooth covariant of order $0$, the regularity (even in the sense
                              of \cite{GS}) of $J_t$ easily follows, and thus $\int_0^t J_s \circ M (ds)$ makes sense.
                               Furthermore, by Ito formula and the q.s.d.e. satisfied by $V_t$, it is simple to verify
                                that $J_t$ satisfies the required map-valued q.s.d.e.   This proves (i).

                                For proving (ii),  our strategy is as in \cite{GS}. We define a family of maps
                                  $\Phi_t$ as follows. First we extend the definition of $J_t$ to make sense
                                   of $J_t(X \ot e(f))$ for $X \in \clb(h_\infty, (h \ot \Gamma_{fr})_\infty)$ where
                                    $\Gamma_{fr}$ is the free Fock space over $\hat{k_0} $ (see \cite{GS}), by setting
                                     $J_t(X \ot e(f))v=(V_t \ot 1_{\Gamma_{fr}})P_{23}(X \ot 1_{\Gamma} )V_t^*(ve(f))$
                                                                   and   then we define for
                                                                   fixed $u,v \in h_0$
                                        and $f,f^\prime$ being bounded continuous $k_0$-valued
                                         functions on $R_+$, $\Phi_t(X,Y)=<J_t(X \ot e(f))u,J_t
                                          (Y \ot e(f^\prime))v>-<ue(f),J_t(X^*Y \ot e(f^\prime))v>$,
                                           for all $X,Y \in \clb(h_\infty,(h \ot \Gamma_{fr})_\infty)$ satisfying
                                            that $X^* \in \clb((h \ot \Gamma_{fr})_\infty,h_\infty)$ (and thus $X^*Y
                                             \in \clb(h_\infty,\infty)$), and in case it is {\bf A7} ,not { $\bf A7^\prime$},
                                               the assumption taken by us, we also require that for any vector $\beta $ in
                                              $\Gamma_f$, $<\beta,X>,<\beta,Y> \in \cla_\infty$. With this definition,
                                               we now fix $x,y \in
                                              \cla_0$ too. We identify $\hat{k_0}^n$ as canonically embedded subspace
                                               of $\Gamma_{fr}$ and it is clear that for $w \in \hat{k_0}^n$,
                                                $\Theta^n(x)_w$ is a smooth map with its adjoint being smooth too
                                                 (which follows from the explicit structure of $\Theta(.)=A(. \ot 1)B$, where
                                                  $A,B$ are smooth covariant, hence have smooth adjoints).
                                                   Thus, for any $n$, and $w,w^\prime \in \hat{k_0}^n$, $\Phi_t(\Theta^n(x)_w,
                                                    \Theta^n(y)_{w^\prime})$ is well-defined and furthermore one can easily verfiy
                                                     relation (3.18) of \cite{GS} with $X=\Theta^n(x)_w,Y=\Theta^n(y)_{w^\prime}$ by using
                                                      Lemma  \ref{thetaalg}.
                                                        We can now iterate this relation arbitrarily many times, and by noting
                                                         the estimate (\ref{basicest}) of Theorem \ref{thest},
                                                          (and also by using the fact that $\| \Phi_t(X,Y) \| \leq
                                                             \{  \| (X \ot 1)V_t^*(ve(f))\|_{2,0}\|(Y \ot 1)V_t^*(ue(f^\prime))\|_{2,0}+
                                                              \|ve(f)\|_{2,0}\| (X^*Y \ot 1)(V_t^*(ue(f^\prime)))\|_{2,0} \},
                                                          $ and $V_t^*$
                                                            is  a smooth covariant map of order $0$, bound $\leq 1$) conclude that $\Phi_t(x,y)=0$
                                                             for $x,y \in \cla_0$. This proves the weak homomorphism property of $j_t$. Since
                                                        $ j_t(x)$ is by the very definition in terms of $V_t$ is a bounded map for all $x \in \bar{\cla}$
                                                          and $\| j_t(x)\| \leq \|x \|_{\infty,0}$, the strong homomorphism property follows.
                                                          The covariance and other properties of $j_t$ as in (ii) of the statement of the present theorem are
                                                           straightforward to see.

                                                           Finally to  prove (iii), first of all we show that $j_t(\bar{\cla}) \subseteq
                                                            \bar{\cla} \ot \clb(\Gamma)$.   For this, we construct iteratively
                                                              $J_t^{(n)}$, by setting $J_t^{(0)}(x \ot e(f))v=(x \ot e(f))v$, and $J_t^{(n+1)}
                                                                              =\int_0^t J_s^{(n)}\circ M_\Theta (ds)$ for $n \geq 0$, and furthermore
                                                                               define  $J^\prime_t(x \ot e(f))v=\sum_n J_t^{(n)}(x \ot e(f))v$, for all
                                                $x,v$ such that the above sum is convergent in the Hilbert space sense.
                                                 One can  verify that for $x \in \cla_0, v \in h_0$,
                                                  $J_t^\prime(x \ot e(f))v$ exists and satisfies the same q.s.d.e. as $J_t$ with the same initial condition, hence
                                                   by the standard iteration argument used to prove the  uniqueness of solution of q.s.d.e., it follows that
                                                   $J^\prime_t(x \ot e(f))v$ equals $J_t(x \ot e(f))v$. Now,
                                                   $(a \ot 1)J_t^{(n)}(x \ot e(f))v=J_t^{(n)}((x \ot e(f))av$ for $x \in \cla_0, v \in h_0$
                                                    and $a \in \bar{\cla}^\prime$ of the form $\clj y \clj$ for some $y \in \cla_0$, where
                                                     $\clj$ is the anti-unitary operator of Tomita-Takesaki theory mentioned before.
                                                     Clearly such an $a$ maps $h_\infty$ into $h_\infty$ and also $\clj \cla_0 \clj$ is
                                                      strongly dense in $\bar{\cla}^\prime$. From this we obtain that $(a \ot 1)
                                                       J_t(x \ot e(f))v=J_t(x \ot e(f))av$ for $x,a,v$ as before. Note that we needed $x \in \cla_0,v \in h_0$
                                                        for showing the summability of $J_t^{(n)}(x \ot
                                                        e(f))v$.Thus
                                                        $j_t(x)$ commutes with all $(a \ot 1), a \in \clj \cla_0 \clj$; and since $j_t(x)$
                                                         is bounded operator, the same holds for all $a \in \bar{\cla}^\prime$. This proves that
                                                         $j_t(\cla_0) \subseteq \bar{\cla} \ot \clb(\Gamma)$, and then due to the normality and
                                                          boundedness of the map $x \mapsto j_t(x)$, the same thing will follow for all $x$.
                                                              Now, take $x \in \cla_\infty$.
                                                              Since $V_t$ is covariant contractive  map,
                                                           we can easily verify that  $<e(f),j_t(x){e(f^\prime)}> \in \bar{\cla}_\infty$ for
                                                            $x \in \cla_\infty, f,f^\prime \in L^2(k_0)_\infty$, and hence by the assumption {\bf A7},
                                                             $<e(f),j_t(x){e(f^\prime)}> \in \cla_\infty$, which completes the proof.
                                                               \qed

    \section{Applications and examples}
    In this section we shall show that it is indeed possible to accommodate many interesting classical and noncommutative
     semigroups in our framework. \\
     \subsection{Classical (commutative) examples}
     First of all we prove that the assumption {\bf A6} regarding crossed product is  valid
      in a typical classical situation.\\
      \bthm
      \label{gpaction}
      Let $X$ be a locally compact separable Hausdorff space with a regular Borel measure $\mu$ on
       it, and let a locally compact group $G$ act on $X$ freely and transitively,
         and the measure $\mu$ is $G$-invariant.
       Then the von Neumann algebra $L^\infty(X,\mu)$ satisfies the assumption {\bf A6}, i.e.
        the crossed product von Neumann algebra $L^\infty(X,\mu)>\!\!\!\lhd G$ is isomorphic with
         the weak closure of the $\ast$-algebra generated by $L^\infty(X,\mu)$ and $u_g$ in $\clb(L^2(X,\mu))$,
       which is in fact the whole of $\clb(L^2(X,\mu))$,   where $u_g$ denotes the unitary representation of $G$ in $L^2(X,\mu)$ induced by the $G$-action
           on $X$.
    \ethm
    {\it Proof:-}\\
    Let $x_0$ be any point of $X$. It is clear  that
          the bijective continuous map $G \ni g \mapsto gx_0 \in X$ is a homeomorphism
          (because $G$ is locally compact and $X$ is Hausdorff, so that any continuous bijection
            from $G$ to $X$ is automatically a homoemorphism).
    Let us denote $L^\infty(X,\mu)$ and $L^2(X,\mu)$ by $\bar{\cla}$ and $h$ respectively. We recall
     that $\bar{\cla} >\!\!\!\lhd G$ can be defined to be the von Neumann algebra generated by
      $f \ot 1, f \in \bar{\cla}$ and $u_g \ot L_g,g \in G$ in
       $\clb(h \ot L^2(G))$ (where $L_g$ is the left regular representation).
        So, the commutant of this von Neumann algebra is the intersection of $\bar{\cla} \ot \clb(L^2(G))$
         and $\{ u_g \ot L_g,g \in G \}^\prime $ (since $\bar{\cla}$ is maximal abelian). But $\bar{\cla} \ot \clb(L^2(G))$ can be identified with
          the direct integral of copies of $\clb(L^2(G))$ over $(X,\mu)$. In this direct integral
           picture, we can view any element $B$ of $\bar{\cla} \ot \clb(L^2(G))$ as a measurable  map
            $B : x \mapsto B(x)\in \clb(L^2(G))$; and then  it is easy to see that $B$ also commutes with
             all $u_g \ot L_g$ if and only if $B(gx)=L_g^*B(x)L_g$ $\forall x,g$.
              Thus, the map $B(.)$ is determined  by the value of
               $B(.)$ at any one point of $X$ (since the action is free and transitive).
                It is now easily seen that
                 $(\bar{\cla} >\!\!\!\lhd G)^\prime$ is isomorphic with $1\ot \clb(L^2(G))$.
                       To verify this, we denote the inverse of the map $g \mapsto gx_0$
                        (which is a homeomorphism as noted earlier) by $\Psi$ and consider
            the unitary $U $ on $h \ot L^2(G) $ given by $(U \phi)(x,g)=\phi(x,\Psi(x)g)$,
             for $\phi \in h \ot L^2(G) \cong L^2(X \times G)$. Clearly, for any $B \in (\bar{\cla} >\!\!\!\lhd G)^\prime,$
              i.e. $B(x)=L_g^*B(x_0)L_g,$ with $g=\Psi(x)$, we have that $UBU^*=1 \ot B(x_0)$, and since
               $B(x_0)$ can be allowed to be an arbitrary element of $\clb(L^2(G))$, we have shown that
                $U(\bar{\cla} >\!\!\!\lhd G)^\prime U^*=(1 \ot \clb(L^2(G)))$, hence $U(\bar{\cla} >\!\!\!\lhd G)U^*=
                 \clb(h) \ot 1$. Now it is enough to prove that the von Neumann algebra generated by
                  $\bar{\cla}$ and $u_g,g \in G$ in $\clb(h)$ is the whole of $\clb(h)$, which is a direct consequence
                   of the imprimitivity theorem.
                             \qed

                             \brmrk
                             Remark :- \\
                            1. The assumption of transitivity  is not so crucial and can be weakened
                              to the assumption that  the bundle $(X,X/G,\pi)$ (where $\pi :X \raro
                               X/G$ is the canonical quotient map) is trivial (which will be true in particular
                                whenever $X/G$ is contractible).
                                   \ermrk

                                   Now, we proceed to give classical examples where our theory works. Let $G$ be a separble
                                    Lie group, equipped with an invariant (w.r.t. the action of $G$ on itself)
                                      Riemannian metric,  $\cla$ be the $C^*$-algebra of continuous functions on $G$
                                    vanishing at
                                    $``\infty"$, and let $G$ act on itself by the left regular action, which is trivially free
                                     and transitive, so that {\bf A6} will be valid.  In case $G$ is
                                       compact, we  check {\bf A7} by using the fact   that
                                        any element of $L^2(G)$  which is almost everywhere differentiable and all the partial
                                         derivatives are again in $L^2$ obviously belongs to the
                                         $C^\infty$-class.For
                                         noncompact $G$, $\bf
                                         A7^\prime$ follows by
                                         Theorem \ref{gpaction}.
                                         The Laplace-Beltrami operator on $G$ commutes with
                                         the isometry group associated with the $G$-invariant Riemannian metric,
                                          and hence
                                          the heat semigroup
                                          generated by it
                                        satisfies the covariance and symmetry (w.r.t. the left Haar measure)
                                       conditions, and also the other assumptions needed for HP dilation theory are satisfied.

                                        As far as the assumptions {\bf A4}, {\bf A5}, {\bf
                                        A6},
                                         {\bf A7} (or { $\bf A7^\prime$}) needed for the EH theory are concerned, we have already
                                         verified {\bf A6},{\bf
                                         A7}
                                         (/{ $\bf A7^\prime$}).
                                           However, one has to verify {\bf A4} and {\bf A5}  case by case, and in many
                                           cases (e.g. compact groups, $R^n$ etc.) these can indeed be verified.

                                      In the context of $R^n$, we now discuss two interesting classes of
                                       q.d.s. First, we consider the expectation   semigroup of a diffusion
                                        process, such that the generator $\cll$  is of the form $\cll(f)=-\sum _{ij}
                                         a_{ij}\partial_i \partial_jf$ for smooth $f$ with compact support, where
                                          $\partial_i $ denotes the $i$-th partial derivative, and $a_{ij}$ are smooth
                                           functions, with the matrix $(( a_{ij} ))$ being pointwise nonsingular and positive
                                            definite, and $x \mapsto (( a_{ij}(x) ))^{-\frac{1}{2}}$ is a smooth bounded function.
                                             Although if $a_{ij}$ are non-constant functions, then the above $\cll$ is not covariant
                                              w.r.t.  the action of translation group, we show now how to choose a different group
                                                acting on $R^n$ such that $\cll$ can be written as $\cll_0+\delta_0$   for some
                                                 covariant 2nd order operator $\cll_0$ and a derivation $\delta_0$. To achieve
                                                  this, we change the canonical Riemannian metric of $R^n$ and equip it with a different
                                                   metric given by $<\partial_i, \partial_j>|_x=b_{ij}(x)$, where $(( b_{ij}(x) )):=
                                                    (( a_{ij}(x) ))^{-1}$. Let us denote by $X$ the Riemannian manifold $R^n$ with this
                                                     new metric and let $G$ be the group of Riemannian isometries of $X$. It is easy
                                                      to verify that if we choose $\cll_0$ to be the generator of the heat semigroup
                                                       on $X$, then $\cll_0$ is $G$-covariant and symmetric
                                                       w.r.t. the Riemannian volume measure, and moreover $\cll$ is indeed the same as
                                                        $\cll_0$ upto some first order operators, which can be written as $\delta_0(.)$ for some
                                                         suitable closed derivation  $\delta_0$ which  generates a 1-parameter automorphism
                                                          group of the underlying function algebra (we omit some technical conditions that may be required
                                                           to ensure the existence of such an automorphism group, but at least for nice enough
                                                            $a_{ij}$ this will be possible). We can now apply our theory on $\cll_0$ to construct dilations,
                                                          and then it is trivial to obtain dilations of $e^{t \cll}$ from the dilation
                                                           of $e^{t \cll_0}$,
                                                           using
                                                           some
                                                           standard
                                                           perturbation
                                                           techniques.
                                                             However, we must point out here that one needs to verify
                                                             {\bf A1}-{\bf A7} case by case. A sufficient condition
                                                             for verifying
                                                               these assumptions is that there is a nice and large enough Lie subgroup of $G$ which acts freely and transitively
                                                                on $X$, (which in particular will imply that $X$ is a Riemannian homogeneous space) and such that $\cla_\infty$ and $\h$ will coincide as sets
                                                                  with those as in the case of $R^n$ with the action of
                                                                  itself.
                                                            For the simple case when $n=1,$ we can show the existence of such a subgroup by
                                                             direct computation which gives us an explicit description of the group of
                                                              isometries of $X$.


                                                                      \subsection{Noncommutative examples}
                                                                      We shall give a class of examples which are closely connected with
                                                                        noncommutative
                                                                        geometry.

                                                                                 \bppsn
                                                                                 \label{ergodic}
                                                                                 Let
                                                                                 $\cla$
                                                                                 be
                                                                                 a
                                                                                 unital
                                                                                 $C^*$
                                                                                 algebra
                                                                                 and
                                                                                 let
                                                                                 $G$
                                                                                 be
                                                                                 a
                                                                                 compact
                                                                                 Lie
                                                                                 group
                                                                                 acting
                                                                                 ergodically
                                                                                 on
                                                                                 $\cla$.
                                                                          Then the assumptions {\bf A4}-{\bf A7} are valid for $\cla$ with the above group action and the unique $G$-invariant normalized
                                                                          trace described in \cite{Alb},\cite{HKr}.
                                                                                 \eppsn
                                                                                 {\it Proof :-}\\

Since $G$ is compact, {\bf A4},{\bf A5} are easy to verify. We shall now prove
that {\bf A6} and {\bf A7}  are also valid. By combining the results of
\cite{Shiga}, \cite{Alb} and \cite{HKr} there is a set of elements
$t^\pi_{ij},\pi \in \hat{G},i=1,...,d_\pi,j=1,...,m_\pi$ of
$\cla$, where $\hat{G}$ is the set of irreducible representation
of $G$, $d_\pi$ is the dimension of the irreducible representation
space denoted by $\pi$, $m_\pi \leq d_\pi$ is a natural number,
such that the followings hold : \\
 (i) The linear span of $\{ t^\pi_{ij} \}$ is norm-dense in
 $\cla$,\\
 (ii) $\{ t^\pi_{ij} \}$ is an orthonormal basis of
 $h=L^2(\cla,\tau)$, \\
 (iii) The action of $u_g$ coincides with the $\pi$-th irreducible
 representation of $G$ on the vector space spanned by
 $t^\pi_{ij},i=1,...,d_\pi$ for each fixed $j$ and $\pi$,\\
 (iv) $\sum_{i=1,...d_\pi} (t^\pi_{ij})^*t^\pi_{ik}=\delta_{jk}
 d_\pi 1$, where $\delta_{jk}$ denotes the Kronecker delta symbol.
Thus, in particular, $\| t^\pi_{ij} \|_{\infty,0} \leq
\sqrt{d_\pi} \forall \pi,i,j$.

Now, we first prove {\bf A6} .  We recall that the crossed product von
Neumann algebra  $\clc:=\bar{\cla}>\!\!\!\lhd G$ is by definition the
von Neumann algebra generated by
   $\{ (t^\pi_{ij} \ot 1),\pi,i,j; (u_g \ot L_g), g \in G \}$ in $L^2(\tau) \ot L^2(G),$
    where $L_g$ is the regular representation of $G$ in $L^2(G)$. Let $\rho$ be the normal
     $\ast$-homomorphism from $\clc$ onto $\{ \bar{\cla},u_g,g \in G \}^{\prime \prime} \subseteq
      \clb(L^2(\tau))$ which satisfies $\rho(t^\pi_{ij} \ot 1)=t^\pi_{ij}$ and $\rho(u_g \ot L_g)=u_g$.
       We have to show that this is an isomorphism, i.e. the kernel of $\rho$ is trivial.
         Clearly, the set of elements of the form $\sum c_{\pi ij} t^\pi_{ij}u_{g_{\pi i j}}$ (finitely many terms),
         with $c_{\pi i j} \in C; g_{\pi ij} \in G$ is dense
         w.r.t. the strong-operator topology in
                      $\{
            \bar{\cla},u_g, g \in G \}^{\prime \prime}$. Similarly, the set of elements of the form
              $\sum c_{\pi i j} (t^\pi_{ij} \ot 1)(u_{g_{\pi i j}} \ot L{g_{\pi i j}}) $ (finitely many terms) will be
              strongly
               dense in $\clc$. Now, let $\cli \equiv \{ X \in \clc : \rho(X)=0 \}$. We need to show that
                                                              $\cli=\{0\}$. Let $X \in \cli$ and let
                                                               $X_p=\sum c_{\pi i j}^{(p)}(t^\pi_{ij} \ot 1)(u_{g^{(p)}_{\pi i j}} \ot L{g^{(p)}_{\pi i j}})  $
                be a net (indexed by $p$) of elements from the above dense algebra such that $X_p$
                 converges strongly to $X$. Hence we have, $\sum
                 |c^{(p)}_{\pi i j }|^2 =\|\rho(X_p)(1)\|_{2,0}^2
                 \raro 0. $ This implies that for any $\phi \in L^2(G),$ $\|X_p(1 \ot \phi)
                 \|^2 =\sum |c^{(p)}_{\pi i j}|^2\|L_{g^{(p)}_{\pi
                 i j}} \phi \|^2 \leq \| \phi \|^2 \sum |c^{(p)}_{\pi i
                 j}|^2\raro 0$, which proves that $X(1 \ot
                 \phi)=0$ for every $X \in \cli$. But since $\cli$
                 is an ideal in $\clc$, this shows that for $a \in
                 \cla$, $X(a \ot \phi)=(X(a \ot 1))(1 \ot
                 \phi)=0,$ and by the fact that $\{ a \ot \phi , a \in \cla, \phi \in
                 L^2(G)\}$ is total in $h \ot L^2(G)$ we conclude
                 that $\cli=\{0\}$.

                 We are now left with the proof of {\bf A7}. Since in
                 this case the trace $\tau$ is finite, we have
                 that $\cla_\infty \subseteq \bar{\cla}_\infty
                 \subseteq \h$. Hence it suffices to prove
                 $\h=\cla_\infty$. Let $\Delta_G$ be the Laplacian
                 on the compact Lie group $G$, and $\lambda_\pi,
                 \pi \in \hat{G}$ be its set of eigenvalues, where
                  $\lambda_\pi$ occurs with $d_\pi^2$
                  multiplicity. Let $v \in \h$ be given by an
                  $L^2$-convergent series $v=\sum c_{\pi i j}
                  t^\pi_{ij}$. The assumption that $v \in \h$
                  implies that $\sum |\lambda_\pi |^{2n} |c_{\pi i
                  j}|^2 < \infty$ for every positive integer $n$.
                  Now, using the well-known Weyl asymptotics for
                   the Laplacian on a compact manifold and the fact that $m_\pi \leq d_\pi$ in our case,
                    it is easy
                   to see that for any large enough $n$, $\sum_{\pi,
                   i,j} \frac{d_\pi}{|\lambda_\pi |^{2n}} <
                   \infty.$ Now, since $\| t^{\pi}_{ij}
                   \|_{\infty,0} \leq \sqrt{d_\pi},$ we have,
                   $\sum_{\pi,i,j} |c_{\pi i j} | \| t^\pi_{ij}
                   \|_{\infty,0} \leq \sum |c_{\pi i j}|
                   \sqrt{d_\pi} \leq \left( \sum |c_{\pi i j}|^2
                   |\lambda_\pi |^{2n} \right)^{\frac{1}{2}} \left( \sum
                   \frac{d_\pi}{|\lambda_\pi|^{2n}}
                   \right)^{\frac{1}{2}} < \infty$ for all large
                   enough $n$. This proves that the series $\sum
                   c_{\pi i j} t^{\pi}_{ij}$ converges in the norm
                   of $\cla$, and hence $v \in \cla$. Similar
                   arguments will enable us to prove that indeed
                   $v \in \cla_\infty$, thereby proving that
                   $\h \subseteq \cla_\infty$, hence
                   $\h=\cla_\infty$.

                               \qed

 We now note that since we have shown in the above proof that
 $\h=\cla_\infty$, and since in the present case the finiteness of
 the trace implies that the Frechet topology of $\cla_\infty$ is
 stronger than that of $\h$, it is clear that the assumption {\bf
 A2}
 implies {\bf A3}. Thus, we have the following nice sufficient condition
 to have EH dilation in this case :\\
 \bppsn
 With the set-up of the Proposition \ref{ergodic}, any covariant, symmetric, conservative  q.d.s. $T_t$ on $\cla$
  such that the domain of its norm-generator contains $\cla_\infty$ in it, admits HP and  EH type dilations.
  \eppsn

                                                                         \subsection{Connection with Arveson-Powers Index Theory}
                                                                         In this final subsection, we would like to apply our results to obtain the numerical
                                                                          index as defined by Arveson and Powers \cite{Po},\cite{Ar} and studied by many other authors.
                                                                           Let us use the notation of the section on HP dilation theory.  In the notation
                                        of the lemma \ref{conservative}, $\tilde{T_t}$ is a covariant q.d.s. on $\clb(h)$ which extends $T_t$. Furthermore,
                                      we can choose $k_0$ in a ``minimal" way, in the sense that there does not exist any
                                          nonzero vector $f \in k_0$ satisfying $<f,\tilde{R}>=0$. To show this, let us denote by $k^\prime$ the
                                           set of all $f \in k_0$ such that $<f,\tilde{R}>=0$. It is easy to check that this set is
                                            a closed subspace of $k_0$, and $(x \ot 1)\tilde{R}\xi \in (h \ot k^\prime)^\perp$ for all $x \in \clb(h),
                                             \xi \in h$. Thus we can replace $k_0$ by $k_0 \ominus k^\prime$ to achieve the required minimality.
                                            For simplicity, let us assume that the trace $\tau$ is finite, so that $1 \in h=L^2(\tau)$.
                                             Assume w.l.g. that $k_0$ has been chosen in a minimal way as described above, and $\{ e_i \}$
                                              be an o.n.b. of $k_0$ consisting of ``smooth" vectors as in the section on HP dilation. Let
                                               $\tilde{R}_i=<e_i,\tilde{R}>$. Note that by construction, $\tilde{R}(1)=0$. We now prove that
                                                the result by Bhat \cite{BhAMS} on the minimality of dilation can be extended to the
                                                 present situation.
                                                 \bthm
                                                 The dilation $\tilde{j}_t$ of $\tilde{T_t}$, given by $\tilde{j}_t(x)=U_t(x \ot 1_\Gamma)U_t^*$,
                                                  (where $U_t$ is the unitary process constructed in the section on HP dilation) is minimal
                                                   in the sense of Bhat and Parthasarathy \cite{BP}, i.e. $\{ \tilde{j}_{t_1}(x_1)...\tilde{j}_{t_n}(x_n)\xi ; x_1,...x_n
                                                    \in \clb(h), \xi \in h; t_1 \geq t_2 \geq ... t_n \geq 0 \}$ is total in $h \ot \Gamma$.
                                                     Thus, the numerical index of $\tilde{T}_t$ in the sense of Arveson is the dimension of
                                                      $k_0$.
                                                      \ethm
                                                      {\it Proof :-}\\
                                                      We first verify the following condition which is sort of $L^2$-independence of the
                                                       family of operators $\{I, \tilde{R}_i;i=1,2,... \}$ :\\
                                                       If $c_0,c_1,...$ is an $l^2$-sequence of complex numbers such that $(c_0I+\sum_i
                                                        c_i \tilde{R}_i)\xi=0 \forall \xi \in h_\infty$, then $c_i=0 \forall i=0,1,2,...$.\\
                                                        To prove this fact, first note that since $1 \in h_\infty$ and $\tilde{R}_i(1)=0 \forall i=1,2,...$
                                                         clearly $c_0$ must be $0$. Now, consider the vector $f \in k_0$ given by $f=\sum_{i\geq 1}
                                                          \bar{c}_i e_i$ (which is well-defined as $c_i$ is $l^2$), and note that the condition
                                                           $\sum c_i \tilde{R}_i=0$ on $h_\infty$ implies that $<f,\tilde{R}>=0$ on $h_\infty$, and
                                                            since $h_\infty$ is a core for $\tilde{R}$, we have that $<f, \tilde{R}>=0$, which implies that
                                                             $f=0$ by our assumption on $k_0$.

                                                             The rest of the proof will be exactly the same as that of \cite{BhAMS}  by Bhat.
                                                             We  note that the boundedness of the coefficients
                                                               (which was assumed by Bhat) is not really used in the
                                                               proof,
                                                               since
                                                               $\h$-vectors
                                                               are
                                                               used.
                                                               \qed

{\bf Acknowledgement :}\\
 The first author (D.G.)  would like to thank Av. Humboldt Foundation for
 a research fellowship during November 2000 to October 2001 when a major part of this work was
  carried out,
  Prof. S. Albeverio for hospitality at the   Inst. F{\"u}r Angew. Math. (Bonn) and
  Indian Stat. Inst. (Kolkata)  for a visiting position for months.
   Furthermore, he thanks P. S.
  Chakraborty, Dr. J. M. Lindsay and Dr. S. J. Wills for many fruitful
  discussions. The second author (K.B.S.) acknowledges the support of the Indo-French
   Centre for Advanced Research. \\


\begin{thebibliography}{AFL}
\bibitem{AFL} L. Accardi, A. Frigerio and J. T.  Lewis,
\emph{Quantum stochastic processes}, Publ.\ Res.\ Inst.\ Math.\
Sci.\ {\bf 18}(no. 1)(1982), 97-133.\\
\bibitem{AK} L. Accardi and S. Kozyrev, \emph{On the structure of
 Markov flows}, to appear in Chaos, Solitons and
 Fractals(2000).\\
\bibitem{Alb} S. Albeverio, R. Hoegh-Krohn, \emph{Ergodic actions by compact groups on
$C^* $-algebras}, Math.\ Z.\ {\bf 174}(no. 1) (1980),  1--17.\\
\bibitem{Ar} W. Arveson,\emph{ The index of a quantum dynamical semigroup},
 J.\ Funct.\ Anal.\ {\bf 146}(no. 2) (1997),  557--588.\\
\bibitem{BhAMS} B. V. R. Bhat, \emph{Cocycles of CCR flows}, Mem.\ Amer.\ Math.\ Soc.\
{\bf 149} (2001), no. 709.\\
\bibitem{BP} B. V. R. Bhat and K. R. Parthasarathy, \emph{ Markov dilations of noncommutative dynamical
 semigroups and a quantum boundary theory}, Ann.\ Inst.\ Henri Poincar\'{e}, Probabilit\'{e}s et
 Statistiqu\'{e}s ({\bf 31}), no. 4 (1995), 601-651. \\
\bibitem{CE} E. Christensen and D. E. Evans, \emph{ Cohomology of operator algebras and quantum dynamical
 semigroups}, J.\ London Math.\ Soc.\ ({\bf 20})(1979), 358-368. \\
\bibitem{CiS} F. Cipriani and J-L. Sauvageot, \emph{Derivations as
Square Roots of Dirichlet Forms}, preprint, Dipartimento di
Matematica, Politenico di Milano (2001).\\
\bibitem{CGS} P. S. Chakraborty, D. Goswami and K. B. Sinha, \emph{A
covariant quantum stochastic dilation theory}, Stochastics in
finite and infinite dimensions,  Trends Math., Birkh{\"a}user
Boston, Boston, MA (2001), 89-99.\\
\bibitem{Dav} E. B. Davies, \emph{Quantum dynamical semigroups and neutron diffusion equation},
Rep.\ Math.\
 Phys. {\bf (11)}(1977), 169-189. \\
\bibitem{Fag} F. Fagnola,  \emph{On quantum stochastic differential equations with unbounded
 coefficients}, Probab.\ Th.\ Rel.\ Fields, ({\bf 86})(1990),
 501-516.\\
 \bibitem{FaS}F. Fagnola and K. B. Sinha, \emph{Quantum flows with unbounded structure maps and finite degrees of freedom.},
  J.\ London Math.\ Soc.\  (2){\bf 48} (no. 3) (1993),
  537--551.\\
 \bibitem{Gar} L. Garding, \emph {Note on continuous representations of
Lie groups}, Proc.\ Nat.\ Acad.\ Sci.\ U.S.A. {\bf 33}(1947),
331-332.\\
\bibitem{GS} D. Goswami and K. B. Sinha , \emph{Hilbert modules and stochastic dilation of a quantum dynamical
 semigroup on a von Neumann algebra}, Commun.\ Math.\ Phys.\
 {\bf 205}, no. 2(1999), 377-403.\\
 \bibitem{GSP} D. Goswami, A. Pal and K. B. Sinha,
 \emph{Stochastic dilation of a quantum dynamical semigroup on a
 separable unital $C^*$ algebra}, Inf.\ Dim.\ Anal.\ Quan.\ Prob.\
 Rel.\ Topics {\bf 3}, no. 1 (2000), 177-184.\\
 \bibitem{HKr}R. Hoegh-Krohn,  M. B. Landstad and E. Stormer,
\emph{Compact ergodic groups of automorphisms} Ann.\ of Math.\ (2)
{\bf 114} ( no. 1)(1981), 75--86.\\
 \bibitem{HP} R. L. Hudson and K. R. Parthasarathy, \emph{ Quantum Ito's Formula and
 Stochastic Evolutions},
 Commun.\  Math.\ Phys.\ ({\bf 93})(1984), 301-323.\\
\bibitem{MoS} A. Mohari and K. B. Sinha, \emph{ Stochastic dilations of minimal quantum dynamical
 semigroup}, Proc.\ Indian Acad.\ Sc.\ ( Math.\ Sc.\ ), {\bf 103} (3)(1992),
 159-173.\\
\bibitem{Nel} E. Nelson, \emph{ Analytic vectors}, Ann.\ of
Math.\  (2){\bf 70}, 1959, 572-615.\\
\bibitem{Par} K. R. Parthasarathy,  \emph{``An Introduction to Quantum Stochastic Calculus"},
 Monographs in Mathematics, Birkh\"{a}user Verlag, Bessel
 (1992).\\
 \bibitem{Po}R. T. Powers,
\emph{An index theory for semigroups of $ \ast$-endomorphisms of
${\cal B}({\cal H})$ and type ${\rm II}_1$ factors}, Canad.\ J.\
Math.\ {\bf 40}(no. 1) (1988),  86--114.\\
\bibitem{Sav} J-L. Sauvageot, \emph{Tangent bimodule and locality
for dissipative operators on $C^*$-algebras}, Quantum Prob.\ and
Applications IV, Lecture Notes in Math.\ {\bf 1396}(1989),
322-338.\\
\bibitem{Shiga} K. Shiga,\emph{ Representations of a compact group on a Banach
space}, J.\ Math.\ Soc.\ Japan {\bf 7} (1955), 224--248.\\


\end{thebibliography}
\end{document}